# Microscopic analysis of the energy, momentum and spin distributions in a surface plasmon-polariton wave

A. Y. Bekshaev[1], O. V. Angelsky[2,3], J. Zheng[2], S. G. Hanson[4], C. Yu. Zenkova[3]

[1]Physics Research Institute, I.I. Mechnikov National University, Odessa, Ukraine
[2]Research Institute of Zhejiang University – Taizhou, Taizhou, China
[3]Chernivtsi National University, Chernivtsi, Ukraine
[4]DTU Fotonik, Department of Photonics Engineering, DK-4000 Roskilde, Denmark
[1]bekshaev@onu.edu.ua
[2]dbzj@netease.com

## Abstract

We analyze the electromagnetic field near a plane interface between a conductive and a dielectric media, under conditions supporting surface plasmon-polariton (SPP) propagation. The conductive medium is described by the hydrodynamic electron-gas model that enables a consistent analysis of the field-induced variations of the electron density and velocity at the interface and its nearest vicinity. The distributions of electromagnetic dynamical characteristics: energy, energy flow, spin and momentum are calculated analytically and illustrated numerically, employing silver-vacuum interface as an example. A set of the "field" and material contributions to the energy, spin and momentum are explicitly identified and classified with respect to their physical origins and properties, and the orbital (canonical) and spin (Belinfante) momentum constituents are separately examined. In this context, a procedure for the spin-orbital momentum decomposition in the presence of free charges is proposed and substantiated. The microscopic results agree with the known phenomenological data but additionally show specific nanoscale structures in the near-interface behavior of the SPP energy and momentum, which can be deliberately created, controlled and used in nanotechnology applications.

## 1. Introduction

Localized light fields attract significant attention due to their impressive abilities within optical nano-probing, precise optical manipulation and optical data processing. Especially, the surface plasmon-polariton (SPP) waves emerging near the interface between dielectric and conductive media are intensively investigated in connection with nano- and micro-optical means for light guiding, switching and controlling, which is crucial for further microminiaturizing of optical information devices and systems [1–6]. Many stimulating and promising applications are related with unique dynamical properties of the SPP waves such as transverse spin and momentum [7–11], special spin-momentum locking [11,12], nonreciprocity and unidirectional propagation [11–14].

The exclusive properties as well as the very existence of the SPP waves are directly associated with the dispersion (frequency dependence of the main electric and magnetic parameters) of both contacting media. Together with the strong spatial inhomogeneity of the SPP-supporting systems, this poses challenges in the theoretical description of the SPP field, especially its dynamical characteristics (DCs): energy, energy flow, momentum, angular momentum and their derivatives. The main problems relating the proper DC description in such "complex" media were recently resolved in a series of works [17–23]. This approach is, in fact, a generalization of the known Brillouin's method for the energy description in lossless dispersive media, which is based on the special dispersive corrections of their permittivities and permeabilities [24,25]. In application to the field momentum and angular momentum, it is closely related to the canonical (spin-orbital) momentum decomposition [26–28]: the dispersion-caused corrections are applied not to the usual Poynting momentum density but to its orbital (canonical) and spin (Belinfante) parts separately.

When being applied to the SPP field, this methodology offers a physically meaningful and consistent description of its momentum and angular momentum with their dependence on the media's properties and the radiation frequency [20–23]. Importantly, this phenomenological approach is justified by the microscopic analysis [20–22] based on the simplest lossless Drude model [2,5] for the electron gas in the metal. However, the Drude model entails some contradictory conclusions relating the SPP field spatial behavior in the near-interface region: the discontinuity of the transverse electric field and non-zero electron velocity at the metal surface lead to non-physical "singular" values of the SPP-induced electron density and to non-realistic data for the plasmonic field enhancement [29].

On the other hand, the microscopic justification of the phenomenological results for the spin and orbital SPP-momentum constituents raises a new problem in the electromagnetic DC theory. Actually, the existing results on the spin-orbital momentum decomposition essentially employ the dual symmetry [30] expressed by the "electric-magnetic democracy" [27] of the Maxwell equations without free charges and currents. In the microscopic electron-gas analysis, the presence of charges and currents destroys this symmetry, making the form and the very existence of a consistent algorithm for the momentum canonical decomposition unclear.

As we will see, both difficulties mentioned in the above paragraphs require a more detailed investigation of the DC spatial distributions with involvement of the electrons' quantum properties, and in the simplest way this can be performed on the ground of the hydrodynamic "jellium" plasma model [29,31–33]. The hydrodynamic model has widely been used in the SPP theory; in particular, some steps were also undertaken in connection with the SPP momentum decomposition [22] but the systematic DC analysis was not accomplished. The solution of this problem is one of the present paper's aims.

In further Sections we present a comprehensive hydrodynamic microscopic description of the SPP field with the main attention paid to its DCs. The first objective is to justify the phenomenological results, including those for the spin-orbital (canonical) momentum decomposition; actually, this task was partly realized [20,22] but now we seek a more reliable record of the microscopic hydrodynamic corrections including those that are usually neglected in the phenomenological picture but are important for the physical consistency. In this process, we perform a second objective: a consecutive identification and classification of the meaningful "blocks" which compose the SPP momentum. This classification includes the "field" and material contributions, differing by their origin and physical nature, as well as their orbital (canonical) and spin parts. And finally, the obtained analytic results are used for the study of the DC properties, especially their spatial details associated with the hydrodynamic model and localized in the near-interface region.

The presentation starts in Section 2 with reproducing the known solutions for the field and material characteristics within the hydrodynamic model framework. This part is intended for completeness and convenience; besides, the known results are presented in the forms suitable for subsequent comments and references. In Section 3 these solutions are applied for deriving the refined expressions of the field DCs. Especially, subsection 3.2 illustrates the calculation of the field-induced material momentum following the general scheme of Ref. [16] and finally presents the SPP momentum as a sum of the "field" and material contributions of different origins; subsection 3.3 describes the general algorithm for the spin-orbital momentum decomposition and details of its application to separate "field" and material contributions. The peculiar features of the SPP angular momentum (which is of the spin nature and has an "extraordinary" transverse direction [7–10]) are analyzed in subsection 3.4. In Section 3, the consideration is concentrated on the DC spatial distributions whereas the radiation frequency (the SPP wavelength) is kept constant; in contrast, Section 4 discusses the phenomena owing to the variable radiation frequency.

The meaning and peculiar features of the new hydrodynamic-based results are disclosed via the systematic juxtaposition with the usual phenomenological expressions for the SPP field and its DCs. To this end, the important known data, summarized and presented in the forms suitable for comparative discussions, are collected in the Supplemental Document. General significance of the main results, their possible manifestations, applications and ways of further development are briefly outlined in Section 5.

## 2. SPP field parameters based on the microscopic hydrodynamic model.

### 2.1. General description of the SPP-supporting system

We consider a standard system supporting the SPP propagation [3–5] (Fig. 1a). Two homogeneous media are separated by the plane interface $x = 0$; medium 1 ($x > 0$) is dielectric with permittivity $\varepsilon_1 > 0$, medium 2 ($x < 0$) is conductive (a metal) characterized by $\varepsilon_2$; both media are supposed to be non-magnetic ($\mu_1 = \mu_2 = 1$). The electromagnetic field is monochromatic, and instantaneous values of the electric and magnetic vectors are represented via the phasors: $\mathbf{E}(t) = \mathrm{Re}\left(\mathbf{E}e^{-i\omega t}\right)$, $\mathbf{H}(t) = \mathrm{Re}\left(\mathbf{H}e^{-i\omega t}\right)$ (from now on, sans-serif symbols denote the real-valued time-dependent quantities while the Roman symbols represent complex phasors). For the SPP physics, the frequency-dependent permittivity $\varepsilon_2 = \varepsilon_2(\omega)$ is crucial; the medium 1 is supposed to be non-dispersive. For simplicity and in compliance with the approach of [20,21], we consider a lossless system, i.e. all the permittivities and permeabilities are real.

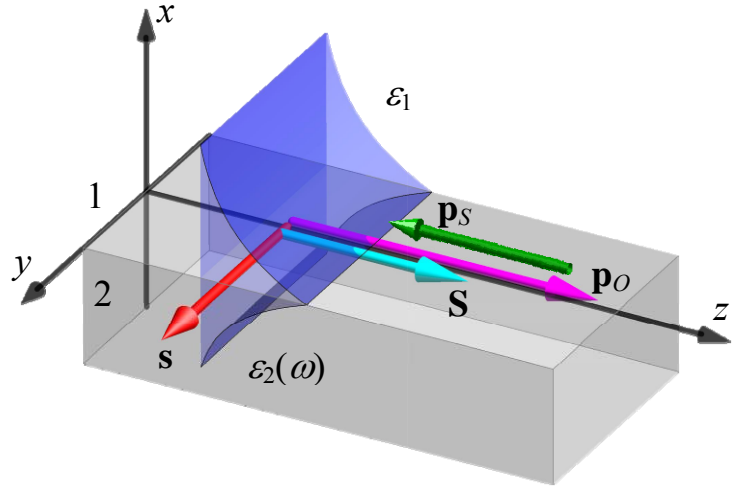

Fig. 1. Geometrical configuration of a system supporting the SPP propagation; the arrows show: (cyan) the energy flow (36), (37), (S8), (S10); (magenta) orbital momentum (62), (S18), (S19); (red) spin (71), (S13), (S15); (green) volume part of the spin momentum (63), (S21).

To start the consideration, we first outline the usual (phenomenological) approach when the materials' constitutive parameters $\varepsilon_1$, $\varepsilon_2 = \varepsilon_2(\omega)$ are postulated prior to the analysis. Then, in each medium, the Maxwell equations hold [24,25]

$$\nabla \mathbf{H} = 0, \quad \mathbf{H} = \frac{1}{ik}\nabla \times \mathbf{E}, \quad \nabla \mathbf{E} = 0, \quad \mathbf{E} = -\frac{1}{ik\varepsilon}\nabla \times \mathbf{H} \tag{1}$$

where $k = \omega/c$, $c$ is the speed of light in vacuum, and the Gaussian system of units is used. In the system of Fig. 1, SPP waves are described by TM solutions of the Maxwell equations [2], and the boundary conditions require that at $x = 0$

$$E_{z1} = E_{z2}; \quad H_{y1} = H_{y2}; \quad \varepsilon_1 E_{x1} = \varepsilon_2 E_{x2}. \tag{2}$$

As a result, the electric and magnetic vectors of the SPP field are obtained in the form [2–7]

$$\mathbf{E} = \frac{A}{\varepsilon_1}\left(\mathbf{x} - i\frac{\kappa_1}{k_s}\mathbf{z}\right)\exp\left(ik_s z - \kappa_1 x\right), \quad \mathbf{H} = \mathbf{y}A\frac{k}{k_s}\exp\left(ik_s z - \kappa_1 x\right), \quad (x > 0); \tag{3}$$

$$\mathbf{E} = \frac{A}{\varepsilon_2}\left(\mathbf{x} + i\frac{\kappa_2}{k_s}\mathbf{z}\right)\exp\left(ik_s z + \kappa_2 x\right), \quad \mathbf{H} = \mathbf{y}A\frac{k}{k_s}\exp\left(ik_s z + \kappa_2 x\right), \quad (x < 0). \tag{4}$$

Here $\mathbf{x}$, $\mathbf{y}$ and $\mathbf{z}$ are the unit vectors of the coordinate axes (see Fig. 1), $A$ is the coordinate-independent normalization constant, and

$$\kappa_{1,2}^2 = k_s^2 - k^2 \varepsilon_{1,2}, \quad k_s^2 = \frac{\varepsilon_1 \varepsilon_2}{\varepsilon_1 + \varepsilon_2}k^2, \quad \frac{\kappa_1}{\kappa_2} = -\frac{\varepsilon_1}{\varepsilon_2}. \tag{5}$$

First Eq. (5) expresses the general relation between the evanescent-wave decay length, the vacuum wavenumber $k$ and SPP wavenumber $k_s$; the second one is the SPP dispersion relation, and the third expresses the useful equality that follows from the first and second ones. As is seen, a propagating SPP mode only exists when $\varepsilon_1$ and $\varepsilon_2$ have opposite signs, and, in view of the supposition $\varepsilon_1 > 0$, $\varepsilon_2$ should be negative and satisfy the condition [4,5,12]

$$\varepsilon_2 < -\varepsilon_1. \tag{6}$$

We suppose that the medium 2 can be characterized as an ideal metal described by the lossless Drude model [2] for which

$$\varepsilon_2 = 1 - \eta, \quad \eta = \frac{\omega_p^2}{\omega^2} \tag{7}$$

where $\omega_p$ is the volume plasmon frequency,

$$\omega_p^2 = \frac{4\pi n_0 e^2}{m}, \tag{8}$$

$e$ is the electron charge ($e < 0$), $m$ is its mass, and the condition (6) means that $\omega < \omega_p/\sqrt{2}$ and $\eta > 2$. The known phenomenological expressions for main DCs of the SPP [2,5,23], following from Eqs. (1) – (8), are presented in the Supplemental Document (Section S1, Eqs. (S1) – (S24)), for the convenience of references and comparison.

Now we address the improvements and refinements to the SPP field description which can be achieved based on the microscopic model of the electron plasma in the conductive medium. These improvements do not affect the field in the dielectric medium 1, and for this reason, all the results below are only applicable to the half-space $x < 0$ (if other is not explicitly indicated).

## 2.2. SPP field and the motion of charges

Following to Refs. [29,31–33] we consider the metal as a combination of the stable uniform "background" charge of ions and the movable electron gas treated within the Bloch hydrodynamic jellium model. In this model, the electrons' motion is characterized by the time-dependent velocity $\mathbf{v}(t)$, and their number density is presented as $n_0 + n(t)$ where the equilibrium uniform density $n_0$ neutralizes the positive charge of ions, and $n(t) \ll n_0$; again, all the relevant quantities can be presented via the complex phasors: $\mathbf{v}(t) = \text{Re}\left(\mathbf{v}e^{-i\omega t}\right)$, $n(t) = \text{Re}\left(ne^{-i\omega t}\right)$. In this case, the first pair of the Maxwell equations (1) remains unchanged but the second pair is modified by the presumption $\varepsilon = 1$ and by explicit presence of the electrons' charge and current:

$$\nabla \mathbf{E} = 4\pi n e, \quad \mathbf{E} = \frac{i}{k}\nabla \times \mathbf{H} - i\frac{4\pi n_0 e}{\omega}\mathbf{v}. \tag{9}$$

The electrons' velocity $\mathbf{v}$ and density $n$ obey the hydrodynamic equation [31]

$$-i\omega n_0 m\mathbf{v} = n_0 e\mathbf{E} - m\beta^2 \nabla n \tag{10}$$

where the quantum statistical effects are included via the coefficient $\beta^2 = (3/5)v_F^2$, involving the Fermi velocity $v_F$. Due to the lossless approximation accepted everywhere in this paper, the electron gas "viscosity" [29] responsible for the damping effects is here omitted (some possible consequences of the SPP energy dissipation are briefly discussed in Section 5). Boundary conditions are also modified and now they include the electron velocity: instead of (2), at $x = 0$ we require [29,31]

$$E_{z1} = E_{z2}; \quad H_{y1} = H_{y2}; \quad \varepsilon_1 E_{x1} = E_{x2}; \quad v_x = 0. \tag{11}$$

The solution of Eqs. (1), (9), (10) and (11) gives a modified expression for the electric field in the metal, instead of the first Eq. (4):

$$\mathbf{E} = \frac{A}{\varepsilon_2}\left[\left(-\eta e^{\gamma x} + e^{\kappa_2 x}\right)\mathbf{x} + i\left(-\eta\frac{k_s}{\gamma}e^{\gamma x} + \frac{\kappa_2}{k_s}e^{\kappa_2 x}\right)\mathbf{z}\right]\exp\left(ik_s z\right); \tag{12}$$

the other results of (3), (4) are still valid. In Eq. (12), $\varepsilon_2$ denotes the expression (7); however, now it is not postulated phenomenologically but derived from Eqs. (1) and (9) − (11). An additional parameter $\gamma$ is introduced that obeys the equation

$$\gamma^2 = k_s^2 - \frac{\omega^2 \varepsilon_2}{\beta^2} \tag{13}$$

(note that according to Eq. (6), $\varepsilon_2 < 0$; this form is very similar to the standard evanescent-wave relations expressed by the 1[st] Eq. (5) (only the replacement $c \rightarrow \beta$ differentiates it from $\kappa_2$). Together with the additional relationship [31,33]

$$\eta\frac{k_s^2}{\gamma} = \kappa_2 + \kappa_1\frac{\varepsilon_2}{\varepsilon_1}, \tag{14}$$

there are four equations for $k_s$, $\kappa_1$, $\kappa_2$, and $\gamma$, which, in particular, "contain" the dispersion relation uniting $k_s$ and $k$. Generally, in the hydrodynamic model, the dispersion relation has no explicit form (like 2[nd] Eq. (5)) but can be calculated numerically, and the result is presented in Fig. 2 (blue curve).

Further results of the model (9), (10) concern the electron gas characteristics:

$$\mathbf{v} = \frac{A}{\varepsilon_2}\frac{e}{m\omega}\left[i\left(-e^{\gamma x} + e^{\kappa_2 x}\right)\mathbf{x} + \left(\frac{k_s}{\gamma}e^{\gamma x} - \frac{\kappa_2}{k_s}e^{\kappa_2 x}\right)\mathbf{z}\right]\exp\left(ik_s z\right), \tag{15}$$

$$n = -\frac{A}{4\pi e}\frac{\eta}{\varepsilon_2}\left(\gamma - \frac{k_s^2}{\gamma}\right)e^{\gamma x}\exp\left(ik_s z\right). \tag{16}$$

Eqs. (12), (15) and (16) constitute the ground for further studies which will include analytical formulas and numerical calculations. For illustrations, we suppose the medium 1 to be vacuum, while the medium 2 parameters are chosen from the electron gas model in silver [35], which gives the main input numerical data:

$$\varepsilon_1 = 1, \quad \omega_p = 7.73 \cdot 10^{15}\ \text{s}^{-1}, \quad v_F = 1.39 \cdot 10^8\ \text{cm/s}, \quad \beta^2 = 1.18 \cdot 10^{16}\ \text{cm}^2/\text{s}^2 \tag{17}$$

(here $\omega_p$ is determined via the Drude-model approximation for Ag [23,34]). In this case, the SPP cutoff frequency equals to $\omega_p\left/\sqrt{1 + \varepsilon_1}\right. = 5.46 \cdot 10^{15}\ \text{s}^{-1}$, and the frequency-dependent behavior of the SPP-field parameters is illustrated by Fig. 2.

The main tasks of this study are the analysis of the solutions (12), (15), (16) and their consequences for the SPP DCs, with special attention to their spatial distributions. For the detailed numerical illustrations, we select the conditions of point A of the dispersion curve in Fig. 2 which corresponds to the radiation frequency $\omega = 4.09 \cdot 10^{15}\ \mathrm{s}^{-1}$ and the vacuum wavelength $\lambda = 2\pi/k = 0.4609\ \mu\mathrm{m}$. Under these conditions, the metal "effective permeability" $\varepsilon_2 = 1 - \eta$, SPP wavelength $\lambda_s = 2\pi/k_s$ and the parameter $\gamma$ become

$$\varepsilon_2 = -2.57, \quad \lambda_s = 0.3663\ \mu\mathrm{m}, \quad \gamma = 6.04 \cdot 10^7\ \mathrm{cm}^{-1}. \tag{18}$$

In all numerical calculations we suppose that $A = 1\ (\mathrm{g/cm})^{1/2}\mathrm{s}^{-1} = 1\ \mathrm{statV/cm} = 3\cdot10^4\ \mathrm{V/m}$ in Eqs. (3), (4) and (12), (15), (16). Accordingly, Fig. 3 shows the relative dimensionless amplitudes of the electric and magnetic fields, electrons' velocity and density, normalized with respect to $A = E_x(0)$.

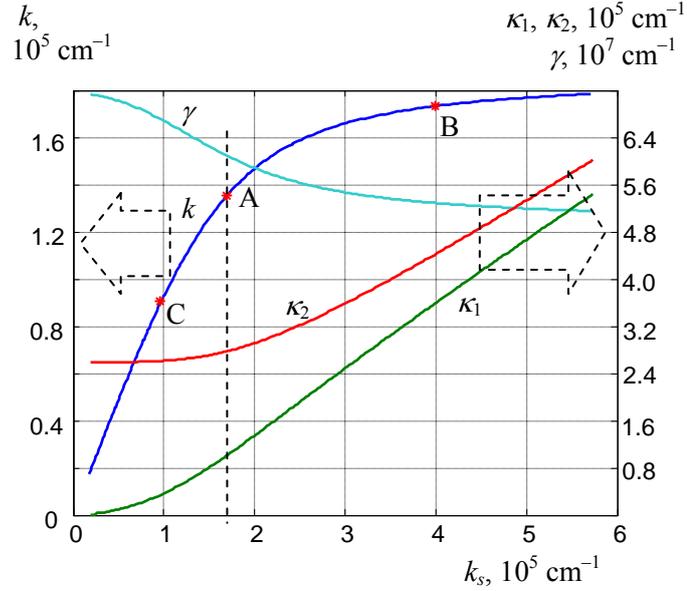

Fig. 2. Frequency dependence of the SPP field characteristics calculated via (13), (14) and (5) for the parameter values (17): (blue) dependence $k(k_s) = \omega(k_s)/c$ (dispersion curve, left vertical scale); (green) $\kappa_1(\omega)$; (red) $\kappa_2(\omega)$; (cyan) $\gamma$ (13) (right vertical scale). Asterisks denote the points on the dispersion curve for which the field and DCs are analyzed in more detail: conditions of point A are applied in Figs. 3 – 8, points B and C are taken for comparison in Fig. 9.

All curves in Fig. 3a (except the brown one for $en/\kappa_2$) contain large segments of smooth exponential off-interface decay determined by factors $\exp(\kappa_2 x)$ $(x < 0)$ and $\exp(-\kappa_1 x)$ $(x > 0)$, which in Fig. 3a look as straight lines. These smooth segments characterize spatial distributions dictated by the phenomenological equations (3), (4); from now on, such contributions and the corresponding terms in the equations will be referred to as the "volume" ones. In the phenomenological description, the "volume" behavior is "in charge" up to the interface (see the blue dashed line for $E_x$ as an illustration). But due to the microscopic refinements, just near the interface, the curves' shapes change rapidly, which is caused by the near-surface (NS) terms of Eqs. (12) – (15) proportional to $\exp(\gamma x)$. The corresponding decay length $1/\gamma$ is impressively smaller than the "volume" counterparts:

$$1/\gamma = 0.165\ \mathrm{nm}, \quad 1/\kappa_2 = 36\ \mathrm{nm}, \quad 1/\kappa_1 = 96\ \mathrm{nm}. \tag{19}$$

Note that all the field components are continuous at $x = 0$ in agreement with the conditions (11) and the 1st Eq. (17).

The NS contributions are absent in curves for $H_y$ and negligible for $E_z$ and $v_z$ (due to $\gamma$ in the denominators of corresponding terms in (12) and (15)), but the rapid NS variations explicitly enable the continuity of $E_x$ and zero value of $v_x$ at the metal surface required by the boundary conditions (11). An especially important fact is that the transverse electric field $E_x$ vanishes at a certain point $x_e$ near the interface (see Fig. 3b; under the conditions (17), (18), $x_e = -0.21$ nm). This fact induces additional peculiar features in the spatial behavior of various field characteristics, as will be shown below (see Figs. 4, 6, 7).

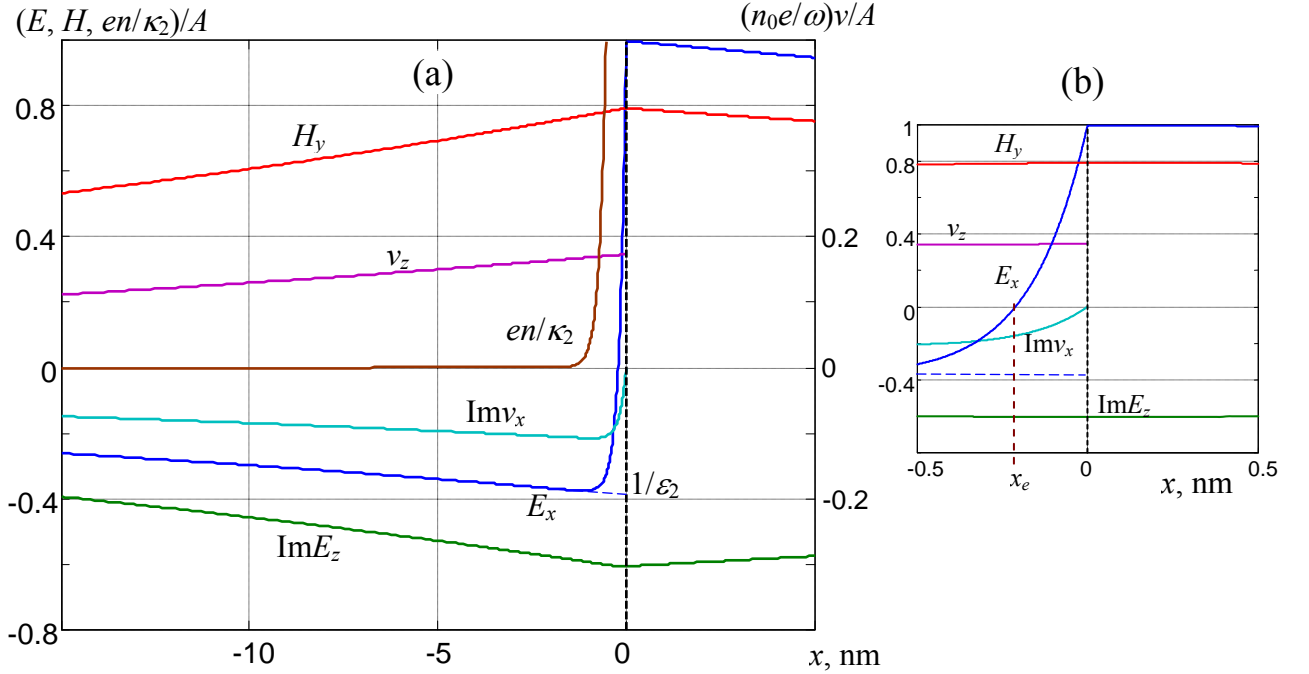

Fig. 3. Amplitudes of (blue, green) electric (3), (12) and (red) magnetic (3), (4) field components, (brown) normalized electron density (16) and (cyan, magenta, right vertical scale) normalized electron velocity components (15) calculated for the conditions (17), (18) (the SPP frequency $\omega = 4.09 \cdot 10^{15}$ s$^{-1}$). The interface plane is marked by the vertical dashed line, the panel (b) shows the NS region of (a) in a magnified horizontal scale.

In agreement with (16), the curve for the SPP-induced charge density $en$ consists completely of the NS segment (the "volume" value of the non-equilibrium electron concentration is zero). Approaching the interface, the concentration amplitude grows so rapidly that the brown curve goes beyond the scale of Fig. 3a and cannot be seen in Fig. 3b; nevertheless, the condition $n \ll n_0$ accepted in the model of (9), (10) is normally not broken: the relative amplitude of the charge-density oscillations at $x = 0$ is

$$\frac{n(0)}{n_0} = -\frac{A}{4\pi e n_0} \frac{\eta}{\varepsilon_2} \left( \gamma - \frac{k_s^2}{\gamma} \right) = \frac{e}{m} \frac{A}{\beta^2 \gamma}. \qquad (20)$$

This equals to $0.739 \cdot 10^{-6} A$ if $A$ is measured in statV/cm, or $2.22 \cdot 10^{-4} A$ if $A$ is expressed in V/cm; that is, even in rather high fields, $n(0)/n_0 \ll 1$ (further discussion see in the concluding Section 5).

For the SPP-induced charge density (16), it is important to know its total (or integral) value obtained by the integration over the whole $x$-range. Such integral values (per unit $y$-width and unit $z$-length) make sense for many characteristics, and the unified notation,

$$\langle ... \rangle = \int\limits_{-\infty}^{\infty} (...) \, dx, \qquad (21)$$

will be used for such quantities throughout the paper (if necessary, the data for $x > 0$ can be taken from Eqs. (3) and from the Section S1 of the Supplemental Document). In application to Eq. (16), $n(x > 0) = 0$ by the definition, and the operation (21) means

$$\langle en \rangle = -\frac{A}{4\pi}\frac{\eta}{\varepsilon_2}\exp\left(ik_s z\right).$$

Now we are in a position to classify the NS terms with respect to their influence on the SPP field characteristics. Eqs. (19) and the curves of Fig. 3a show that, as a rule, the terms of Eqs. (12), (15), (16) with $\gamma$ in denominators ($\gamma^{-1}$-terms) make a minor influence and can be safely discarded when compared with the other summands (in the graphs, their contributions are almost undistinguishable). This is not an exclusive consequence of our choice of the medium 2 and of the "working" frequency (point A in Fig. 2) – the Fermi-velocity data for other metals [35] confirm that, in the wide frequency range, the conditions

$$\beta^2 \to 0, \quad \gamma \to \infty \tag{22}$$

are normally valid. Importantly, upon conditions (22), Eq. (14) reduces to the last equality (5), and $\kappa_1$, $\kappa_2$ and $k_s$ are well described by Eqs. (5); in particular, the dispersion curve of Fig. 2 practically coincides with what follows from the simplified dispersion law of the 2$^{\text{nd}}$ Eq. (5).

Nevertheless, the $\gamma^{-1}$-terms may produce meaningful contributions upon differentiation with respect to $x$; that is why their explicit presence in Eqs. (12) – (16) and in some equations below is crucial for the consistent analysis. On the other hand, the role of NS terms can be "suppressed" by the integration, and, generally, the terms proportional to $\exp(\gamma x)$ make negligible contributions to the integral values (21), except those containing $\gamma$ as a multiplier. As far as (22) holds, such terms can be represented via the delta function:

$$\gamma \exp(\gamma x) \simeq \delta(x). \tag{23}$$

For example, the electron density (16) can be expressed in the form

$$n \simeq -\frac{A}{4\pi e}\frac{\eta}{\varepsilon_2}\exp\left(ik_s z\right)\delta(x) \tag{24}$$

which is actually the limiting form of the electron density dictated by the Drude model [29]. Terms in the form (23), (24) frequently occur in the hydrodynamic description of the medium 2 characteristics when a certain quantity almost vanishes in the volume, but shows an impetuous growth approaching the interface (beside the brown curve in Fig. 3a, see the magenta and green curves in Fig. 5, black and light-green curves in Fig. 7). Due to the delta-shaped behavior, we will call such terms "singular" and use, sometimes, the simplified notation through the delta-function; however, keeping in mind that the boundary values of such terms remain finite, and the delta-function at any moment can be replaced by (23) with the definite peak value $\delta(0) = \gamma$.

## 2.3. Electric polarization of the medium 2

The results (12) – (16) supply a complete characterization of the electron plasma in medium 2 and, particularly, describe its SPP-induced polarization. In the monochromatic field each electron performs an oscillatory motion whose velocity $\mathbf{v}$ is associated with the displacement $(i/\omega)\mathbf{v}$ from its "neutral" position. This displacement produces an effective dipole with a dipole moment per unit volume

$$\mathbf{d} = n_0 (ie/\omega)\mathbf{v}, \tag{25}$$

and Eqs. (12), (15) determine the linear dependence

$$\mathbf{d} = \hat{\alpha}\mathbf{E}. \tag{26}$$

Here $\hat{\alpha}$ is, generally, a diagonal spatially-inhomogeneous tensor with the elements

$$\alpha_{xx}(x) = \alpha \frac{e^{\kappa_2 x} - e^{\gamma x}}{e^{\kappa_2 x} - \eta e^{\gamma x}}, \quad \alpha_{zz}(x) = \alpha \frac{\kappa_2 e^{\kappa_2 x} - \left(k_s^2/\gamma\right)e^{\gamma x}}{\kappa_2 e^{\kappa_2 x} - \eta\left(k_s^2/\gamma\right)e^{\gamma x}}, \tag{27}$$

where

$$\alpha = -\frac{\eta}{4\pi} = \frac{\varepsilon_2 - 1}{4\pi} \tag{28}$$

is merely the phenomenological polarizability of the medium 2 [2,24,25]. The quantities $\alpha_{xx}$ and $\alpha_{zz}$ describe its microscopic refinements and spatial variability: from the "volume" value $\alpha_{xx}(-\infty) = \alpha_{zz}(-\infty) = \alpha$ to the surface values $\alpha_{xx}(0) = 0$, $\alpha_{zz}(0) = \alpha \dfrac{\kappa_2 - \left(k_s^2/\gamma\right)}{\kappa_2 - \eta\left(k_s^2/\gamma\right)}$. Under conditions (22), the spatial dependence of $\alpha_{zz}(x)$ can usually be neglected, but the variations of $\alpha_{xx}(x)$ are crucial for the NS behavior of the SPP characteristics. For example, Eqs. (15) and (27) reflect the non-zero electron velocity at a point $x_e$ where $E_x = 0$: formally, in this point denominator of $\alpha_{xx}(x)$ (27) vanishes, and $\alpha_{xx}(x)$ is singular. As we will see, this singularity is important for the spatial dependence of the field characteristics but it does not complicate their formal analysis because in subsequent equations $\alpha_{xx}$ occurs only in the form of products like $\alpha_{xx}E_x$ where the singularity of $\alpha_{xx}$ is cancelled by the zero of $E_x$.

"Transverse" and "longitudinal" polarizabilities $\alpha_{xx}$, $\alpha_{zz}$ depend on frequency $\omega$. This dependence will be used below, and a useful expression for the derivative can be inferred from (27):

$$\alpha'_{xx} = \frac{d\alpha_{xx}}{d\omega} = \alpha' \frac{e^{2\kappa_2 x} - e^{(\kappa_2 + \gamma)x}}{\left(e^{\kappa_2 x} - \eta e^{\gamma x}\right)^2} + \alpha x \varepsilon_2 \frac{\left(\kappa'_2 - \gamma'\right)e^{(\kappa_2 + \gamma)x}}{\left(e^{\kappa_2 x} - \eta e^{\gamma x}\right)^2}. \tag{29}$$

Here primes denote derivatives with respect to $\omega$, and the relations dictated by Eq. (28),

$$\alpha' = \frac{d\alpha}{d\omega} = \frac{\eta}{2\pi\omega} = \frac{1 - \varepsilon_2}{2\pi\omega} = \frac{1}{4\pi}\varepsilon'_2, \quad \alpha'\eta = \alpha\eta', \tag{30}$$

are used. In Eq. (29), the second summand vanishes at $x = 0$ and rapidly decays at $x < 0$ due to the exponential multiplier, so it merely supplies a minor correction that can safely be omitted in most cases compatible with the condition (22). As for the "longitudinal" polarization $\alpha_{zz}(x)$, as far as its spatial dependence is negligible, the "volume" approximations $\alpha_{zz} = \alpha$, $\alpha'_{zz} = \alpha'$ are relevant.

### 3. Dynamical characteristics of the SPP field.

Now we proceed to calculations of the DCs for the SPP field in the metal layer. In what to come, we distinguish the "field" contributions marked by superscript "$F$", and the material contributions denoted by superscript "$M$". The results for medium 1, obtained phenomenologically (see Section S1), will naturally be treated as "field contributions". Also, we will keep the following classification scheme for the spatially-dependent terms:

- "volume" terms proportional to $\exp(2\kappa_2 x)$ and describing the "smooth" exponential decay in the depth of medium 2;
- NS terms proportional to $\exp(\gamma x)$ describing the rapid off-surface decay and giving negligible contributions upon spatial integration;
- "singular" terms proportional to $\gamma \exp(\gamma x)$ describing the strictly-localized near-surface contributions reducible to the delta-function, see (23);
- $\gamma^{-1}$-terms proportional to negative degrees of $\gamma$. These terms are normally negligible and omitted but sometimes kept in intermediate transformations.

### 3.1. Energy density

In the model of Eqs. (9), (10), the energy density in metal is determined by formula [31,33]:

$$w = \frac{1}{16\pi}\left(|\mathbf{E}|^2 + |\mathbf{H}|^2\right) + \frac{1}{4}mn_0|\mathbf{v}|^2 + \frac{1}{4}\frac{m\beta^2}{n_0}|n|^2 = w^F + w^M \, ; \tag{31}$$

the field contribution is expressed by the first summand, the second and third ones describe the material part. By using Eqs. (12) − (16) and neglecting the $\gamma^{-1}$-terms we obtain

$$w^F = \frac{g}{2\varepsilon_2^2}\left[\left(1 + \varepsilon_2 + \frac{\varepsilon_2^2}{\varepsilon_1} - \frac{\varepsilon_2}{\varepsilon_1}\right)e^{2\kappa_2 x} - 2\eta e^{(\kappa_2+\gamma)x} + \eta^2 e^{2\gamma x}\right], \tag{32}$$

$$w^M = \frac{g\eta}{2\varepsilon_2^2}\left[\left(1 - \frac{\varepsilon_2}{\varepsilon_1}\right)e^{2\kappa_2 x} - 2e^{(\kappa_2+\gamma)x} + \eta e^{2\gamma x}\right], \tag{33}$$

$$w = \frac{g}{\varepsilon_2^2}\left[\frac{\varepsilon_1 - \varepsilon_2 + \varepsilon_2^2}{\varepsilon_1}e^{2\kappa_2 x} - 2\eta e^{(\kappa_2+\gamma)x} + \eta^2 e^{2\gamma x}\right], \quad w(0) = g\frac{\varepsilon_2 + \varepsilon_1\varepsilon_2 - 1}{\varepsilon_1\varepsilon_2}, \tag{34}$$

where $g = |A|^2/(8\pi)$, see Supplemental Document.

First, compare these results to the phenomenological expression (S4). The "field" and material parts are not resolved in (S4) but the (first) "volume" term of (34) expectedly coincides with the phenomenological data. The NS terms contribute to sharp variations near the interface $x = 0$, well seen in Fig. 4. In contrast to Fig. 3, the NS terms produce not only rapid growth or decrease but also an additional "kink" near the point $x_e$ related to the vanishing amplitude $E_x$ (see Fig. 3b). The NS terms make no effect on the integral values determined by the "volume" first terms of (32) − (34) but affect the boundary relations. In particular, the energy density "step" predicted by the phenomenological formula (S5) is illustrated by the vertical segment CD in Figs. 4a, b; the "microscopically corrected" step following from (34),

$$\Delta w = g\frac{1 - \varepsilon_1\varepsilon_2}{\varepsilon_1\varepsilon_2}, \tag{35}$$

characterized by the segment CE, is higher. Interestingly, the "field" part of the energy density (blue curve) is distributed continuously (due to the continuity of the electric and magnetic fields, see (11) and (17)), so the complete energy discontinuity (35) follows from the 2nd and 3rd material terms of Eq. (31), which exists only in the medium 2.

Now consider the energy flow density. In the microscopic approach and in application to the TM SPP field, it is given by equation [33]

$$S = \frac{c}{8\pi}E_x^* H_y + \frac{1}{2}m\beta^2 n^* v_z \, . \tag{36}$$

It contains the material part additional to the field contribution (S7) and expressed by the second summand of (36). According to (13) − (16), this material part can be presented as

$$\frac{g}{\varepsilon_2}\frac{\omega}{\gamma}\eta e^{\gamma x}\left(\frac{k_s}{\gamma}e^{\gamma x} - \frac{\kappa_2}{k_s}e^{\kappa_2 x}\right)$$

which fully consists of $\gamma^{-1}$-terms (see the classification of the 1st paragraph of Section 3), and, due to our convention, can be discarded, whereas the first summand of (36) via the immediate application of second Eq. (4) and Eq. (12) leads to the expression

$$S = \frac{g}{\varepsilon_2}\frac{\omega}{k_s}\left[e^{2\kappa_2 x} - \eta e^{(\kappa_2+\gamma)x}\right]. \tag{37}$$

The first term of the right-hand side coincides with the phenomenological result (S8) but due to the NS summand of (37), the energy flow step at the interface differs from (S9):

$$\Delta S = g\,\frac{\omega}{k_s}\left(\frac{1}{\varepsilon_1}-1\right). \tag{38}$$

In the example of Fig. 4, the "phenomenological" step AB is reduced to $\Delta S = 0$ because the electric and magnetic fields are continuous ($\varepsilon_1 = 1$, see (17) and (38)). Besides, Eq. (37) shows that the energy flow inversion occurs inside the medium 2 rather than exactly at the interface. The "inversion point" is the same point $x_e$ where $E_x$ changes its sign (cf. Figs. 3b and 4b) – quite expectedly in view of the definition (S7), (36). And the general behavior of the microscopic energy flow in the NS region looks rather similar to that of the transverse electric field component. The integral energy flow is still determined by (S10).

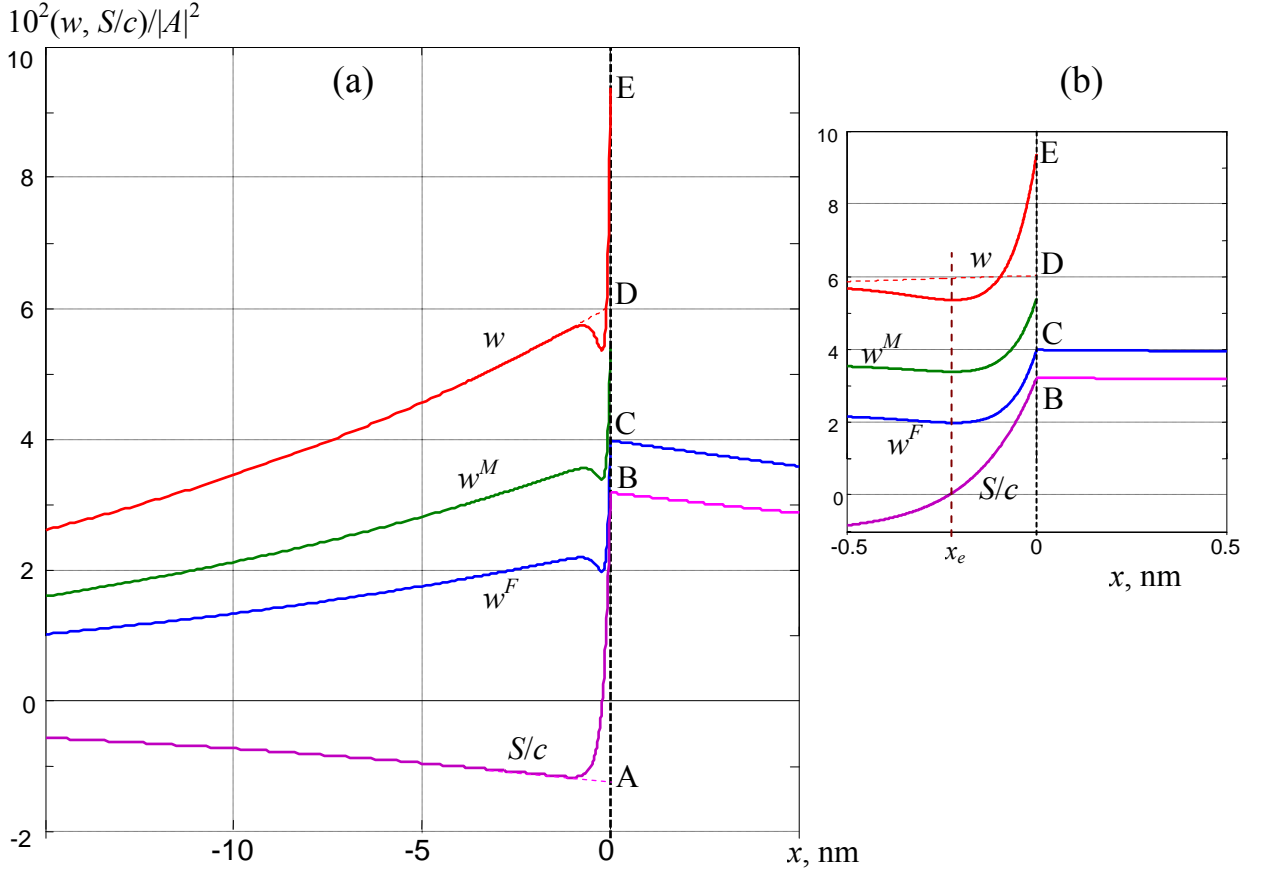

Fig. 4. Energy and energy flow density distributions calculated for the conditions (17), (18): (blue) "field" contribution (32) and, for $x > 0$, (S4); (green) material contribution (33); (red) "complete" energy density (34); (magenta) energy flow density (37) and (S8). Dashed lines show the phenomenological results (without the NS terms in (34) and (37)). D and A: surface values of the phenomenological distributions (S4), (S8) for $x < 0$; C and B: surface values of distributions (S4) and (S8) for $x > 0$. The interface plane is marked by the vertical dashed line, the panel (b) shows the NS region of (a) in a magnified horizontal scale.

### 3.2. Momentum density.

Our next step is the microscopic calculation of the momentum in medium 2. In contrast to the phenomenological analysis (S16) – (S24) [23], which was based on the standard formal procedures, the microscopic approach resembles the process of "building" the "complete" momentum from certain "blocks" of different physical origins.

The first momentum constituent is evidently the "pure-field" contribution expressed by the Poynting vector

$$\mathbf{p}^F = \mathbf{z}p^F = \frac{1}{8\pi c}\operatorname{Re}\left(\mathbf{E}^* \times \mathbf{H}\right), \quad p^F = \frac{1}{8\pi c}E_x^* H_y. \tag{39}$$

In view of (S7), this quantity is closely related to the energy flow density (37), $p^F = c^{-2}S$, and

$$p^F = \frac{g}{c}\frac{k}{\varepsilon_2 k_s}\left[e^{2\kappa_2 x} - \eta e^{(\kappa_2 + \gamma)x}\right]. \tag{40}$$

In principle, its behavior can be understood from the above discussion of $S$ (Fig. 4) but in the "momentum context" it is again illustrated by the blue curves in Fig. 5. Remarkably, the "volume" part of (40) perfectly coincides with the volume part of the phenomenological momentum (S23).

Other "blocks" of the electromagnetic momentum in medium 2 include the material contributions associated with the motion of electrons. To find these contributions, we must consider the force exerted by the field on the medium particles, which is performed in the Supplementary section S2 and dictates that the material contribution consists of two parts. The first one is determined by Eq. (S29) and can be presented as

$$\mathbf{p}_m^M = -\frac{\eta}{8\pi c}\operatorname{Re}\left(\mathbf{E}^* \times \mathbf{H}\right) + \frac{1}{2c}\left(\frac{e\beta^2}{\omega^2}\nabla n \times \mathbf{H}\right) = \mathbf{z}p_m^M, \tag{41}$$

where, in view of Eqs. (13), (16) and (4),

$$p_m^M = -\eta p^F + \frac{g}{c}\eta\frac{k}{k_s}e^{(\kappa_2 + \gamma)x}, \tag{42}$$

and $p^F$ is determined by (39), (40) so that finally

$$p_m^M = -\frac{g}{c}\frac{\eta}{\varepsilon_2}\frac{k}{k_s}\left[e^{2\kappa_2 x} - e^{(\kappa_2 + \gamma)x}\right], \quad p_m^M(0) = 0. \tag{43}$$

Its spatial behavior is visualized by the yellow curves in Fig. 5. Note that the NS term causes it to vanish at the boundary. "Block" (41) – (43) represents the material part of the Poynting momentum proportional to $\eta = 1 - \varepsilon_2$ [16].

The second material "block" of the SPP momentum, which can be called "electric" due to direct connection to the electric field components, follows from (S30) in the form

$$\mathbf{p}_e^M = \mathbf{z}p_e^M, \quad p_e^M = \frac{1}{4}\operatorname{Im}\left(\alpha_{xx}'E_x^*\frac{\partial E_x}{\partial z} + \alpha_{zz}'E_z^*\frac{\partial E_z}{\partial z}\right) = \frac{k_s}{4}\left(\alpha_{xx}'\left|E_x\right|^2 + \alpha_{zz}'\left|E_z\right|^2\right). \tag{44}$$

For the SPP field of Eqs. (4), (12) and employing only the 1st term in the right-hand side of Eq. (29), this can be expressed as

$$p_e^M = \frac{g}{c}\frac{\eta}{\varepsilon_2^2}\frac{k_s}{k}\left[\left(1 - \frac{\varepsilon_2}{\varepsilon_1}\right)e^{2\kappa_2 x} - e^{(\kappa_2 + \gamma)x}\right], \quad p_e^M(0) = -\frac{g}{c}\frac{k_s}{k}\frac{\eta}{\varepsilon_1\varepsilon_2}. \tag{45}$$

This approximation is shown by the orange curves in Fig. 5. For comparison, the results of a direct application of Eq. (44) with numerical differentiation for $\alpha_{xx}'$ and $\alpha_{zz}'$ are presented by the cyan curves: the difference may reach ~10% but it is localized in the very narrow NS region and practically does not affect the general character of the spatial dependence (45).

Yet another material contribution appears due to the rotational motion of electrons. Its calculation starts from the "material rotational spin" [20–22,31]:

$$\mathbf{s}_R^M = \frac{mn_0}{2\omega}\operatorname{Im}\left(\mathbf{v}^* \times \mathbf{v}\right) = \mathbf{y}s_R^M \tag{46}$$

which, via Eq. (15), entails

$$s_R^M = -\frac{2g}{\omega}\frac{\eta}{\varepsilon_2^2}\frac{\kappa_2}{k_s}\left[e^{2\kappa_2 x} - \left(1 + \frac{k_s^2}{\gamma\kappa_2}\right)e^{(\kappa_2+\gamma)x} + \frac{k_s^2}{\gamma\kappa_2}e^{2\gamma x}\right].\tag{47}$$

Hence, the "material rotational spin momentum" follows from the spin momentum definition (S20) in the form $\mathbf{p}_{SR}^M = \mathbf{z}p_{SR}^M$ with

$$\begin{aligned}p_{SR}^M &= \frac{1}{2}\frac{ds_R^M}{dx} = -\frac{g}{\omega}\frac{\eta}{\varepsilon_2^2}\frac{\kappa_2}{k_s}\left[2\kappa_2 e^{2\kappa_2 x} - \left(\kappa_2\frac{\varepsilon_2-\varepsilon_1}{\varepsilon_2}+\gamma\right)e^{(\kappa_2+\gamma)x} - 2\kappa_2\frac{\varepsilon_1}{\varepsilon_2}e^{2\gamma x}\right]\\&\simeq \frac{g}{c}\frac{k_s}{k}\frac{1-\varepsilon_2}{\varepsilon_1\varepsilon_2}\left[2e^{2\kappa_2 x} - \frac{\varepsilon_2-\varepsilon_1}{\varepsilon_2}e^{(\kappa_2+\gamma)x} - 2\frac{\varepsilon_1}{\varepsilon_2}e^{2\gamma x} - \frac{1}{\kappa_2}e^{\kappa_2 x}\delta(x)\right]\end{aligned}\tag{48}$$

where the relation (23) has been used; note that the $\gamma^{-1}$-terms are held in (47) because these make meaningful contributions in (48). The expected boundary value of the "volume" part of $p_{SR}^M$ (given by the first summand in brackets of (48)) is

$$\left(p_{SR}^M\right)_{\mathrm{Vol}}(0) = \frac{2g}{c}\frac{k_s}{k}\frac{1-\varepsilon_2}{\varepsilon_1\varepsilon_2}$$

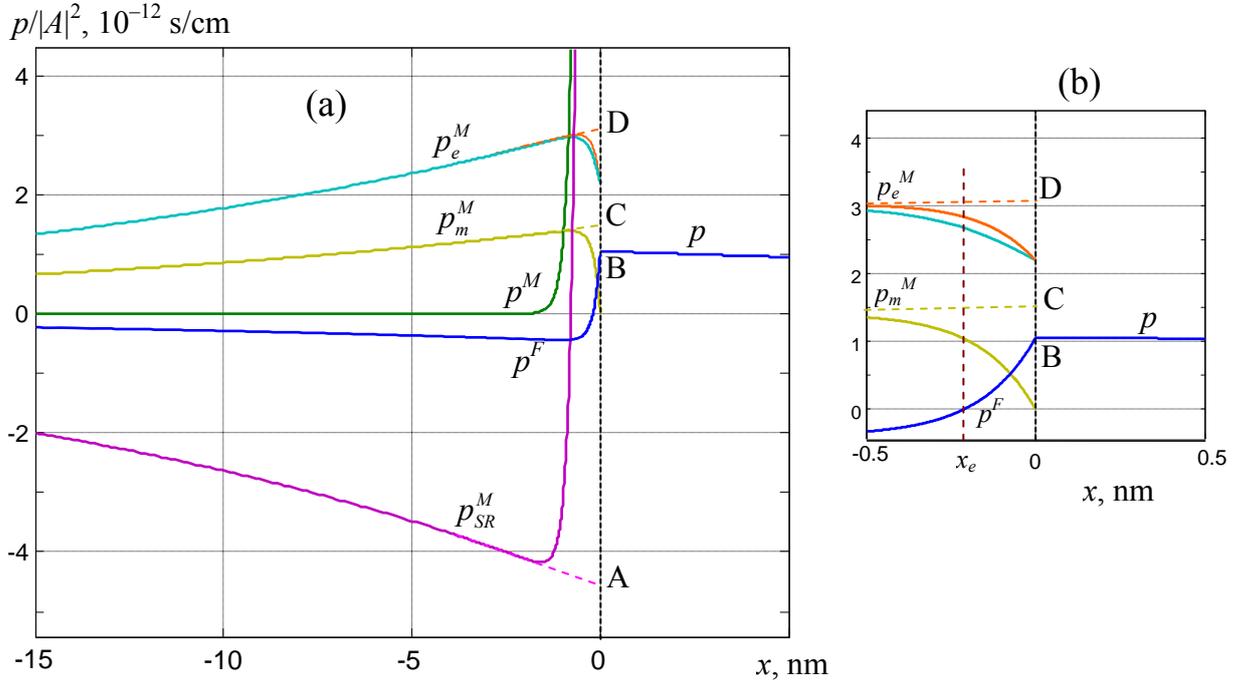

Fig. 5. SPP momentum constituents calculated for the conditions (17), (18): "field" and material contributions. (Blue) "field" contribution (40) (at $x>0$ coincides with the complete momentum (S23)); (yellow) material Poynting momentum (43); (cyan) material "electric" contribution (44); (orange) approximation (45); (magenta) "material rotational spin momentum" (48); (green) complete material momentum (49). Dashed lines show the phenomenological results (without the NS terms in (43), (45) and (48)); D, C and A show the phenomenological distributions for $x<0$ extrapolated to the interface $x=0$ (S4), (S8); B: surface value of $p^F$ (40) coinciding with the surface value of $p$ at $x>0$ (S23). The interface plane is marked by the vertical dashed line, the panel (b) shows the NS region of (a) in a magnified horizontal scale.

(see point A in Fig. 5a). The "regular" NS contributions (2nd and 3rd terms in brackets of (48)) are practically unperceivable on the background of the "singular" delta term; the latter appears due to the fast decay of the spin (47) from its volume behavior (1st summand in parentheses (47)) to zero, caused by the second (NS) summand (see Fig. 8, cyan curve).

Interestingly, the complete material contribution $p^M = p_e^M + p_m^M + p_{SR}^M$ reduces to the "singular" and NS terms (i.e., "volume" material momentum vanishes):

$$p^M \simeq \frac{g}{\omega} \frac{k_s}{\varepsilon_2^2} \eta \left[ e^{(\kappa_2 + \gamma)x} - 2e^{2\gamma x} + \frac{1}{\kappa_1} e^{\kappa_2 x} \delta(x) \right] \qquad (49)$$

which is well illustrated by the green curve in Fig. 5a. As a result, the overall electromagnetic momentum $p$ in the medium 2 (not shown in Fig. 5) appears as a sum of the pure-field contribution (40) (blue curve in Fig. 5) and the "singular" momentum (49) with a near-vertical ramp of the green curve in the interface vicinity:

$$p = p^F + p^M = \frac{g}{c} \frac{k}{\varepsilon_2 k_s} \left[ e^{2\kappa_2 x} - \eta \frac{\varepsilon_2}{\varepsilon_1 + \varepsilon_2} e^{(\kappa_2 + \gamma)x} - 2\eta \frac{\varepsilon_1}{\varepsilon_1 + \varepsilon_2} e^{2\gamma x} + \eta \frac{\kappa_2}{\varepsilon_2 k^2} e^{\kappa_2 x} \delta(x) \right]. \qquad (50)$$

Noteworthy, both the "volume" and "singular" terms perfectly agree with the phenomenological expression (S23), and only the NS terms allow for refinements inspired by the boundary conditions for $E_x$ and $v_x$ (11) in the microscopic theory. This agreement confirms that all "blocks" (40), (43), (45), and (48), which form the SPP momentum in medium 2, are identified correctly, and the microscopic theory complies with the phenomenological approach presented in Supplemental Document.

Additionally, the microscopic analysis discloses the nature of the "singular" spin momentum $p^{surf}$ of (S21) and shows that its "true" boundary value is finite:

$$p(0) \simeq p_M(0) \simeq p_{SR}^M(0) \simeq -\frac{g}{c} \frac{k_s}{k} \frac{\gamma}{\kappa_2} \frac{1 - \varepsilon_2}{\varepsilon_1 \varepsilon_2} \qquad (51)$$

[which equals to $5.05 \cdot 10^{-10}$ g/(cm$^2$·s) with conditions (17), (18)]. In this expression, non-"singular" terms of (48) – (50) are neglected because, due to (19), the right-hand side of Eq. (51) exceeds the other momentum constituents by two orders of magnitude. Because of the high absolute values, only the initial stages of the exponential "ramp" of the green and magenta curves, approaching the interface, are shown in Fig. 5a.

## 3.3. Spin-orbital momentum decomposition

In the phenomenological approach of Section S1 (see Supplemental Document), the spin-orbital (canonical) momentum decomposition is basic for the momentum calculation. In the course of the microscopic approach it is more natural to rely on another meaningful subdivision of the momentum constituents – namely, separation of the material and "field" contributions, which was made in the previous subsection. Now we consider the spin-orbital decomposition in the microscopic framework. To this end, we separately inspect each constituent of the field momentum calculated in subsection 3.2,

$$p = p_e^M + p_{SR}^M + p^F + p_m^M. \qquad (52)$$

We begin with the first terms whose attribution is quite evident. One can easily see that expression (44), upon discarding the NS terms of (29) and taking (30) into account, can be represented as

$$\mathbf{z} p_e^M = \frac{1}{4} \mathbf{z} \, \text{Im} \left( \alpha_{xx}' E_x^* \frac{\partial E_x}{\partial z} + \alpha_{zz}' E_z^* \frac{\partial E_z}{\partial z} \right) \simeq \frac{1}{16\pi} \frac{d\varepsilon_2}{d\omega} \text{Im} \left[ \mathbf{E}^* \cdot (\nabla) \mathbf{E} \right];$$

that is, $p_e^M$ implies the dispersion correction of the orbital momentum (S16) and is therefore its part. On the other hand, $p_{SR}^M$, originating from the spin (46), belongs to the spin momentum "by the definition".

Other terms need a more careful investigation. Regarding the pure-field momentum $p^F$, its decomposition immediately follows from the formal decomposition of the Poynting vector (39), which in view of the Maxwell equations (9) reads [20,22]

$$\mathbf{p}^F = \mathbf{p}_S^F + \mathbf{p}_O^F \tag{53}$$

where

$$\mathbf{p}_S^F = \frac{1}{32\pi\omega}\operatorname{Im}\left[\nabla\times\left(\mathbf{E}^*\times\mathbf{E}\right)+\nabla\times\left(\mathbf{H}^*\times\mathbf{H}\right)\right]-\frac{e}{4\omega}\operatorname{Im}\left(\mathbf{E}^*n\right), \tag{54}$$

$$\mathbf{p}_O^F = \frac{1}{16\pi\omega}\operatorname{Im}\left[\mathbf{E}^*\cdot\left(\nabla\right)\mathbf{E}+\mathbf{H}^*\cdot\left(\nabla\right)\mathbf{H}\right]-\frac{n_0 e}{4\omega c}\operatorname{Im}\left(\mathbf{H}^*\times\mathbf{v}\right). \tag{55}$$

This form differs from the usual spin-orbital decomposition [27,28,30] valid only for fields without free charges and currents. The presence of charges in Eqs. (9) destroys the "electric − magnetic democracy" [27] and produces additional charge-dependent terms (last summands of (54) and (55)) whose attribution is not dictated by general principles and can be ambiguous. The form (54), (55) is heuristic and apparently contradicts to some natural expectations: e.g., the expression (54) leads to a non-zero integral value and thus violates the spin-momentum property (S22). Such controversies emerge because in presence of free charges the Poynting momentum (39) is only a part of the entire electromagnetic momentum. As we will see, they are removed after involvement of other momentum contributions, and the decomposition (53) − (55) is justified, at least for the SPP field of Eqs. (3), (4), (12) − (16).

Now we proceed to further transformations keeping in mind the specific SPP field properties (in particular, that $\mathbf{H}^*\times\mathbf{H}=0$ for the TM modes); as usual, the NS terms proportional to $\gamma^{-1}$ can be omitted. Then the summands of expression (54) obtain the forms

$$\frac{1}{32\pi\omega}\operatorname{Im}\left[\nabla\times\left(\mathbf{E}^*\times\mathbf{E}\right)\right]=-\mathbf{z}\frac{g}{2\omega}\frac{\kappa_2^2}{\varepsilon_2^2 k_s}\left[2e^{2\kappa_2 x}-\eta\frac{\varepsilon_2-\varepsilon_1}{\varepsilon_2}e^{(\gamma+\kappa_2)x}-2\eta^2\frac{\varepsilon_1}{\varepsilon_2}e^{2\gamma x}-\eta\frac{\gamma}{\kappa_2}e^{(\gamma+\kappa_2)x}\right],$$

$$-\frac{e}{4\omega}\operatorname{Im}\left(\mathbf{E}^*n\right)=-\mathbf{z}\frac{g}{2\omega}\frac{\kappa_2^2}{\varepsilon_2^2 k_s}\left[\eta^2\frac{\varepsilon_1}{\varepsilon_2}e^{2\gamma x}+\eta\frac{\gamma}{\kappa_2}e^{(\kappa_2+\gamma)x}\right],$$

which show that the "singular" $\gamma$-proportional terms of both summands mutually cancel. Since the delta-shaped ($\gamma$-proportional) "singularity" of the "field" spin momentum at $x=0$ is hardly interpretable and looks non-physical, such a cancellation testifies that the term $\sim\operatorname{Im}\left(\mathbf{E}^*n\right)$ has been reasonably attributed to the spin part of the decomposition. As a result, an explicit expression for the spin part of the "field" momentum (54) follows in the form

$$\mathbf{p}_S^F = \mathbf{z}p_S^F,$$

$$p_S^F = \frac{g}{2\omega}\frac{\kappa_2^2}{\varepsilon_2^2 k_s}\left[-2e^{2\kappa_2 x}+\eta\frac{\varepsilon_2-\varepsilon_1}{\varepsilon_2}e^{(\kappa_2+\gamma)x}+\eta^2\frac{\varepsilon_1}{\varepsilon_2}e^{2\gamma x}\right],\quad p_S^F(0)=\frac{g}{2\omega}\frac{k_s}{\varepsilon_2}\left(1+\frac{1}{\varepsilon_1}-\varepsilon_2+\frac{\varepsilon_2}{\varepsilon_1}\right) \tag{56}$$

In (55), we modify the term with $\operatorname{Im}\left(\mathbf{H}^*\times\mathbf{v}\right)$. To this purpose we remark that for a TM mode, the Maxwell equations (1), (9) yield

$$\frac{1}{8\pi c}\operatorname{Re}\left[\mathbf{E}^*\times\mathbf{H}\right]=\frac{1}{8\pi\omega}\operatorname{Im}\left[\mathbf{H}^*\cdot\left(\nabla\right)\mathbf{H}\right]-\frac{n_0 e}{2\omega c}\operatorname{Im}\left(\mathbf{H}^*\times\mathbf{v}\right);$$

besides, Eqs. (12), (15), (25), (26) enable us to write

$$\frac{1}{8\pi c}\operatorname{Re}\left[\mathbf{E}^*\times\mathbf{H}\right]=-\frac{1}{8\pi c}\operatorname{Re}\left[-i\frac{e}{\omega}\hat{\alpha}^{-1}\mathbf{v}^*\times\mathbf{H}\right]=-\frac{m\omega}{8\pi ec}\operatorname{Im}\left[\mathbf{H}^*\times\mathbf{v}\right]+\mathbf{z}\frac{g}{c}\frac{k}{k_s}e^{(\kappa_2+\gamma)x}.$$

Therefore,

$$-\frac{n_0 e}{4\omega c} \mathrm{Im}\left(\mathbf{H}^* \times \mathbf{v}\right) = \frac{1}{16\pi\omega} \frac{\eta}{\varepsilon_2} \mathrm{Im}\left[\mathbf{H}^* \cdot (\nabla)\mathbf{H}\right] - \mathbf{z}\frac{g}{2\omega}\frac{k^2}{k_s}\frac{\eta}{\varepsilon_2} e^{(\kappa_2 + \gamma)x}.$$

Note that the "orbital-like" form of this term (cf. Eq. (S16)) confirms its attribution to the orbital part (55) of the decomposition (53). As a result, the "field" orbital momentum (55) acquires the form

$$\mathbf{p}_O^F = \frac{1}{16\pi\omega} \mathrm{Im}\left[\mathbf{E}^* \cdot (\nabla)\mathbf{E} + \frac{1}{\varepsilon_2}\mathbf{H}^* \cdot (\nabla)\mathbf{H}\right] - \mathbf{z}\frac{g}{2\omega}\frac{k^2}{k_s}\frac{\eta}{\varepsilon_2} e^{(\kappa_2 + \gamma)x} = \mathbf{z}p_O^F;$$

$$p_O^F = \frac{g}{2\omega}\frac{k_s}{\varepsilon_2^2}\left(2e^{2\kappa_2 x} - \eta\frac{3\varepsilon_1 + \varepsilon_2}{\varepsilon_1}e^{(\kappa_2 + \gamma)x} + \eta^2 e^{2\gamma x}\right), \quad p_O^F(0) = \frac{g}{2\omega}\frac{k_s}{\varepsilon_2}\left(1 - \frac{1}{\varepsilon_1} + \varepsilon_2 + \frac{\varepsilon_2}{\varepsilon_1}\right). \quad (57)$$

By using Eqs. (5), one can easily verify that the sum of (56) and (57) coincides with (40), and that the "volume" terms coincide with the phenomenological results (S18), (S21).

At last, the decomposition of $p_m^M$ can be performed quite similarly due to relation (42) connecting it to the "field" momentum $p^F$. The only problem that remains is to separate properly the spin and orbital parts of the additional NS summand in (42), which is solved by the natural requirement for both parts to vanish at the interface. Accordingly, $p_m^M = p_{mO}^M + p_{mS}^M$ where

$$p_{mO}^M = -\frac{g}{2\omega}\frac{k_s}{\varepsilon_2^2}\eta\left[2e^{2\kappa_2 x} - \left(2 + \eta^2\right)e^{(\kappa_2 + \gamma)x} + \eta^2 e^{2\gamma x}\right], \quad (58)$$

$$p_{mS}^M = -\frac{g}{2\omega}\frac{k_s}{\varepsilon_2^2}\eta\left[2\frac{\varepsilon_2}{\varepsilon_1}e^{2\kappa_2 x} - \left(2\frac{\varepsilon_2}{\varepsilon_1} - \eta^2\right)e^{(\kappa_2 + \gamma)x} - \eta^2 e^{2\gamma x}\right]. \quad (59)$$

Interrelations between the elements of the spin-orbital decomposition and their spatial distributions are illustrated in Fig. 6 where they are confronted with the corresponding momentum constituents in medium 1 determined by (S18), (S21). One can see that the "volume" orbital and spin constituents are directed oppositely, so the complete momentum appears as a result of their "competition". Notably, in medium 1 the orbital momentum $p_O^F$ is "stronger"; its direction coincides with the complete momentum direction but the absolute value is higher, $\left|p_O^F\right| > \left|p^F\right|$ (in particular, this makes the mechanical action of an evanescent wave stronger than that of a plane wave with the same energy [7,8], see the note below Eq. (S24)). In contrast, in the conductive medium 2 the directions of the complete momenta $p^F$, $p_m^M$ agree with the directions of their spin constituents $p_S^F$, $p_{mS}^M$, and the orbital parts are generally weaker: $\left|p_O^F\right| < \left|p^F\right|$, $\left|p_{mO}^M\right| < \left|p_m^M\right|$. Also, while the directions of the "field" constituents are constant everywhere at $x < 0$ (e.g., $p_O^F > 0$, $p_S^F < 0$), the material contributions $p_{mS}^M$ and $p_{mO}^M$ have sign reversals in the NS region; besides, there exist a narrow segment enclosing $x_e$ (zero point of $E_x$, see Fig. 3 and Fig. 6b), where $p_{mS}^M$ and $p_{mO}^M$ are both positive. The zero-line crossings as well as the sharp "kinks" of the red, cyan and magenta curves in the NS region (well resolved in Fig. 6b) are associated with the sign inversion of $E_x$ and occur near $x_e$. Remarkably, the point, where $p_m^M = p_{mO}^M$, coincides with $x_e$ (this can be immediately inferred from Eqs. (12) and (58), (59) if $\varepsilon_1 = 1$).

Fig. 6 shows that while the phenomenological results ("volume" terms) stipulate a behavior discontinuous at the interface for all the considered quantities, the NS terms restore the continuity of the momentum constituents. Indeed, the blue, green and red curves of the "field" contributions $p_O^F$, $p_S^F$ and $p^F$ are continuous with their counterparts in medium 1 (herewith, the continuity of $p^F$ is "absolute" and follows from the definition (39) and the boundary conditions for $\mathbf{E}$ and $\mathbf{H}$ (11), whereas $p_O^F$ and $p_S^F$ are continuous only in the considered situation $\varepsilon_1 = 1$). The material spin and orbital constituents vanish at $x = 0$, which can also be treated as a sort of continuity because the material contributions in medium 1 are absent.

The data of Fig. 6 illustrate mainly the technical details of the spin-orbital decomposition. In addition, Fig. 7 presents the results of its application to meaningful characteristics of the SPP field. Such characteristics are the complete material orbital momentum

$$p_O^M = p_e^M + p_{mO}^M = \frac{g}{\omega} \frac{k_s}{\varepsilon_2^2} \eta \left[ -\frac{\varepsilon_2}{\varepsilon_1} e^{2\kappa_2 x} + \frac{\eta^2}{2} e^{(\kappa_2 + \gamma)x} - \frac{\eta^2}{2} e^{2\gamma x} \right] \quad (60)$$

and the complete material spin momentum

$$p_S^M = p_{SR}^M + p_{mS}^M = \frac{g}{\omega} \frac{k_s}{\varepsilon_2^2} \eta \left[ \frac{\varepsilon_2}{\varepsilon_1} e^{2\kappa_2 x} + \left(1 - \frac{\eta^2}{2}\right) e^{(\kappa_2 + \gamma)x} + \left(\frac{\eta^2}{2} - 2\right) e^{2\gamma x} + \frac{1}{\kappa_1} e^{\kappa_2 x} \delta(x) \right] \quad (61)$$

which include not only the elements of the formal spin-orbital decomposition (53) – (55) but also the terms $p_e^M$ and $p_{SR}^M$ (see (52), (45), (48)) that were classified at the beginning of this Section based on their immediate physical meaning. Then, by combining the "field" and material contributions, we build the complete spin and orbital momenta of the field,

$$p_O = p_e^M + p_O^F + p_{mO}^M = \frac{g}{\omega} \frac{k_s}{\varepsilon_2} \left[ \left( \frac{1}{\varepsilon_2} - \frac{1}{\varepsilon_1} + \frac{\varepsilon_2}{\varepsilon_1} \right) e^{2\kappa_2 x} + \frac{\eta}{2} \left( \frac{\eta^2 - 3}{\varepsilon_2} - \frac{1}{\varepsilon_1} \right) e^{(\kappa_2 + \gamma)x} + \frac{\eta^2}{2} e^{2\gamma x} \right], \quad (62)$$

$$p_O(0) = \frac{g}{2\omega} \frac{k_s}{\varepsilon_2} \frac{\varepsilon_1 + \varepsilon_2 \varepsilon_1 - 3\eta}{\varepsilon_1};$$

$$p_S = p_{SR}^M + p_S^F + p_{mS}^M$$
$$= \frac{g}{\omega} \frac{k_s}{\varepsilon_2} \left[ \frac{2 - \varepsilon_2}{\varepsilon_1} e^{2\kappa_2 x} + \eta \left( \frac{1}{2\varepsilon_1} + \frac{\eta^2}{2} + \frac{\varepsilon_1 - \varepsilon_2}{\varepsilon_1 \varepsilon_2} \right) e^{(\kappa_2 + \gamma)x} + \eta \left( \frac{\eta^2}{2} - \frac{2}{\varepsilon_2} \right) e^{2\gamma x} + \frac{\eta}{\varepsilon_2 \kappa_1} e^{\kappa_2 x} \delta(x) \right]. \quad (63)$$

It is expressions (62) and (63) that should be immediately juxtaposed with the phenomenological results of Eqs. (S18) and (S21) but even Eqs. (60), (61) 'per se' permit to draw some physically consistent conclusions. In particular, they tell that the "volume" parts of $p_O^M$ and $p_S^M$ have the same absolute values but are directed oppositely (the "volume" parts of the light-green and cyan curves in Fig. 7a are symmetric with respect to the horizontal axis). That is why in the complete material momentum $p^M$ they exactly compensate each other, due to which Eq. (49) only contains the NS and "singular" terms. Remarkably, despite that the "volume" contributions of $p_O^M$ and $p_S^M$ mutually cancel, the integral material contributions to the orbital and spin momenta over the whole medium 2 are equal:

$$\left\langle p_O^M \right\rangle_2 = \left\langle p_S^M \right\rangle_2 = \frac{g}{2\omega} \frac{\eta k_s}{\varepsilon_2^2 \kappa_1} \quad (64)$$

($\langle \ldots \rangle_2$ means the integration like (21) but limited to the $x < 0$ half-space). This happens due to the "singular" term in $p_S^M$ (61) that "acts" oppositely to the "volume" one and produces a twice higher integral contribution.

Further, cyan and magenta curves in Fig. 7a demonstrate, in full agreement with Eqs. (59) and (60), that the "volume" parts of $p_O^M$ and $p_{mS}^M$ coincide, but in the NS region these quantities behave quite differently (see Fig. 7b). Generally, in the NS region all the curves show characteristic "kinks" originating from the transverse-electric-field sign inversion and whose nature was already discussed in connection to Figs. 4 – 6. Exclusions are the black and light-green curves presenting the material $p_S^M$ and the complete $p_S$ spin momenta. In these dependencies, the dominating role belongs to the "singular" $\delta$-shaped term (originating from the material "rotational" spin momentum $p_{SR}^M$, see the magenta curve in Fig. 5a) whose sharp growth masks all fine details. To make them visible, the "residual" spin momentum

$$p_S - p_{SR}^M = p_S^F + p_{mS}^M = \frac{g}{\omega}\frac{k_s}{\varepsilon_2}\left[\frac{\varepsilon_2}{\varepsilon_1}e^{2\kappa_2 x} + \frac{\eta}{2}\left(\frac{1}{\varepsilon_1}+1+\eta\right)e^{(\kappa_2+\gamma)x}-\frac{\eta^2}{2}e^{2\gamma x}\right], \quad (65)$$

which is the complete spin momentum with exclusion of the "material rotational" constituent (48), has been calculated, and the result is illustrated by the yellow curve in Fig. 7, where the "kink" is distinctly seen.

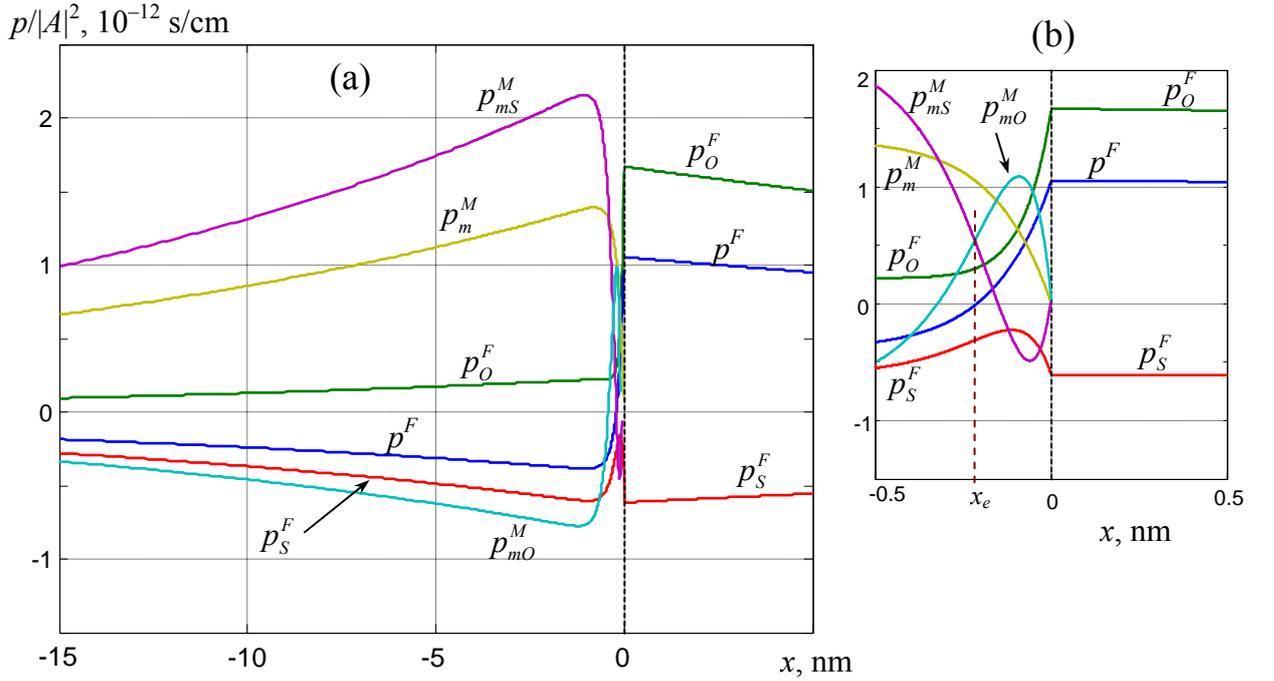

Fig. 6. SPP momentum constituents calculated for the conditions (17), (18): Spin-orbital decomposition of the "field" (40) and material (43) momentum. (Blue) complete "field" momentum (40) and (S23) (for $x > 0$); (green) orbital "field" momentum (57) and (S18) (for $x > 0$); (red) spin "field" momentum (56) and (S21) (for $x > 0$); (cyan) orbital material contribution (58); (magenta) spin material contribution (59); (yellow) material Poynting contribution (43). Blue and yellow curves are the same as in Fig. 5. The interface plane is marked by the vertical dashed line, the panel (b) shows the NS region of (a) in a magnified horizontal scale.

Note that only the illustrative, artificial quantity $p_S - p_{SR}^M$ shows a continuity with the spin momentum in medium 1; this is an occasional property related to the continuity of $p_S^F$ and vanishing $p_{mS}^M$ at $x = 0$ (see Fig. 6). All other momentum constituents discussed in Fig. 7, including the complete orbital and spin momenta $p_O$ (62) and $p_S$ (63) experience "jumps" at $x = 0$, which can be attributed to the fact that these are formed as combinations of the "field" and material parts, whereas in medium 1 only the "field" contributions exist. Noteworthy, the NS terms destroy the perfect proportionality between the orbital momentum $p_O$ (62) and energy density $w$ (34) distributions that are characteristic for the phenomenological approach [23] (cf. Eqs. (S17) and (S4), (S18)): their difference in the NS region is shown by the grey and brown curves in Fig. 7b. However, the boundary values $p_O(0)$ and $w(0)$ in Fig. 7b visually coincide (this is an artifact of the supposition $\varepsilon_1 = 1$ (17)).

Finally, one can see that the "volume" parts of $p_S$ and $p_O$ perfectly agree with the phenomenological expressions (S18) and (S21) as well as the "singular" term of (63) agrees with $p^{\text{surf}}$ in (S21). Accordingly, all integral (in the sense of (21)) properties of the phenomenological expressions hold for the microscopic model. In particular, the microscopically defined orbital momentum obeys integral relation (S19), and the integral spin momentum is equal to zero (as it is dictated by Eq. (S22) and required by the general theory [27,28] despite that it was derived with the use of intermediate spin-momentum constituents based on Eq. (54) which apparently violates this requirement. As to the "surface" $\delta$-term of (S21), the microscopic analysis reveals its "true" finite boundary value (see (51)) which, due to (19), is two orders higher than other momentum contributions described by Figs. 5 – 7 and is thus not shown in the figures. Again discarding the relatively small non-"singular" terms as it was made in (51), we find that the result (51) is also correct for $p_S(0) \simeq p_S^M(0)$.

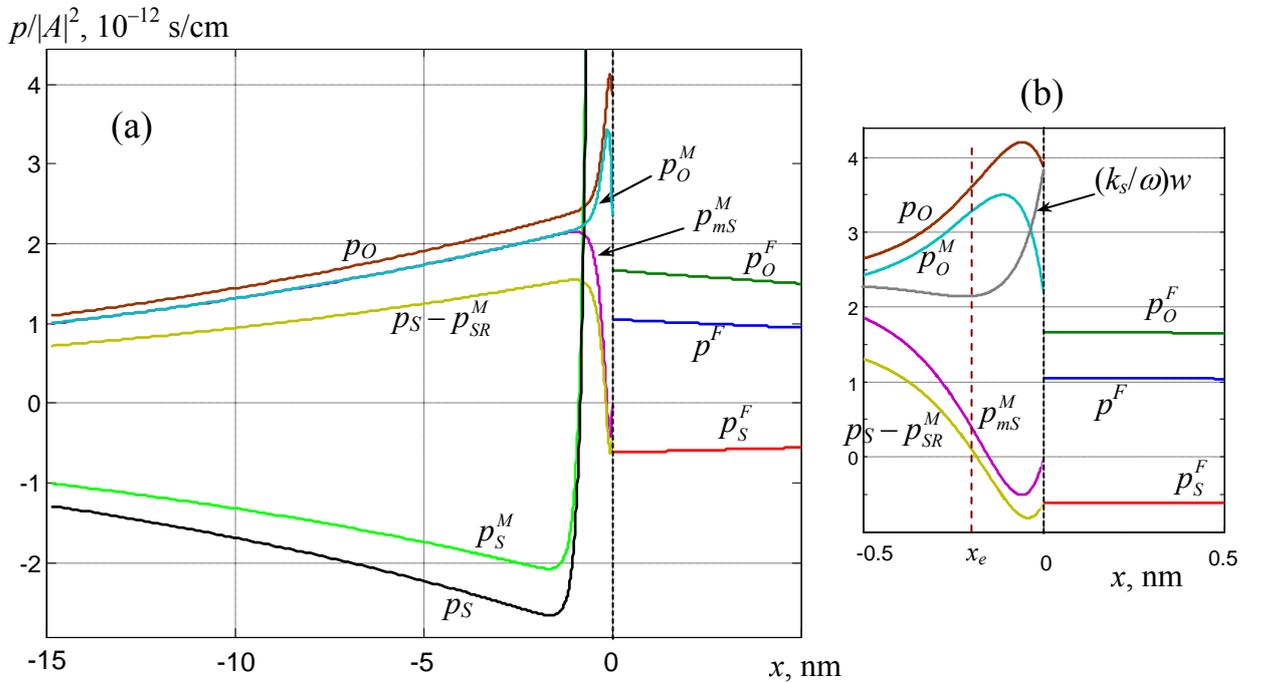

Fig. 7. SPP momentum constituents calculated for the conditions (17), (18): Complete spin and orbital contributions and their material ingredients. (Blue) complete "field" momentum (S23) (for $x > 0$); (green) orbital "field" momentum (S18) (for $x > 0$); (red) spin "field" momentum (S21) (for $x > 0$); (brown) complete orbital momentum (62); (black) complete spin momentum (63); (cyan) orbital material momenum (60); (light-green) spin material momentum (61); (magenta) spin material contribution (59) (the same as in Fig. 6); (yellow) "residual" spin momentum (65); (grey) energy density (34) multiplied by ($k_s/\omega$) according to (S17). The interface plane is marked by the vertical dashed line, the panel (b) shows the NS region of (a) in a magnified horizontal scale.

## 3.4. Spin density of the SPP

A natural way for the microscopic spin calculation is formally putting together different contributions associated with the rotational motion of the field vectors and field-driven material particles [31,20,22]. Formally, this way requires a detailed inventory of the field-induced rotations in the medium, similarly to what was performed upon considering the momentum contributions in subsection 3.2. One of the spin contributions has already been mentioned: it is the material spin associated with the rotational motion of electrons (46), (47), which was employed as an auxiliary item for the momentum analysis. Here we reproduce this contribution in the final form with omitted $\gamma^{-1}$-terms:

$$s_R^M = -\frac{2g}{\omega}\frac{\eta}{\varepsilon_2^2}\frac{\kappa_2}{k_s}\left[e^{2\kappa_2 x} - e^{(\kappa_2+\gamma)x}\right]. \tag{66}$$

This "rotational" material spin is illustrated by the cyan curves in Fig. 8. It satisfies the natural condition $s_R^M(0) = 0$: the very rapid variation from the expected "volume" value A to zero is caused by the NS term. This term is also responsible for the "singular" surface component of the momentum $p_{SR}^M$ in (48).

Another spin contribution is the "field" spin which formally follows from Eq. (S12) applied to the microscopic TM field in medium 2 with $\varepsilon = \mu = 1$ and obeying Eqs. (9). However, the "formal" spin definition (S12) was obtained for the fields without free charges, and its non-critical use may miss some medium-associated corrections that were already applied for the momentum calculations above. To avoid this, the corresponding spin contributions will be calculated based on the relation (S20) between the spin and spin momentum due to which, for the SPP geometry, any **z**-oriented spin momentum "block" $p_S$ is related to the associated **y**-directed spin contribution $s$:

$$p_S = \frac{1}{2}\frac{ds}{dx}. \tag{67}$$

Hence, the spin expression can be recovered via integration of $p_S(x)$ with the natural boundary condition $s(-\infty) = 0$. Then, based on the results for $p_S^F$ (56) and $p_{mS}^M$ (59) we obtain the corresponding "field" and material "blocks" of the spin density in medium 2 (terms proportional to $\gamma^{-1}$ are omitted as usual):

$$s^F = -\frac{g}{\omega}\frac{\kappa_2}{\varepsilon_2^2 k_s}e^{2\kappa_2 x}, \quad s_m^M = -\eta s^F = \frac{g}{\omega}\eta\frac{\kappa_2}{\varepsilon_2^2 k_s}e^{2\kappa_2 x}, \tag{68}$$

cf. the blue and green curves in Fig. 8.

It is instructive to compare this result with the "formal spin" derived from the formal spin definition (S12)

$$s^{Formal} = \frac{1}{16\pi\omega}\mathbf{y}\cdot\text{Im}\left(\mathbf{E}^*\times\mathbf{E}\right) = s^F + \frac{g}{\omega}\frac{\kappa_2}{k_s}\frac{\eta}{\varepsilon_2^2}e^{(\kappa_2+\gamma)x} \tag{69}$$

(light-green curves in Fig. 8). First to note, $s^{Formal}$ reduces to the "field" spin $s^F$ (however, modified by the NS term): the specific material contribution $s_m^M$ does not appear in the formal way of the spin calculation. Second, both expressions (68) contain no NS terms and show the smooth "volume" variation up to the interface (Fig. 8) while the spin $s^{Formal}$ rapidly changes near the interface, reversing the sign together with $E_x$ in point $x_e$, cf. Fig. 3b. The NS term in (69), responsible for this difference, appears because the term $\sim \text{Im}\left(\mathbf{E}^*n\right)$ of (54), taken into account in (56) and, consequently, in (68), is omitted in $s^{Formal}$ (69). As a result, $s^F$ is discontinuous at $x = 0$ while $s^{Formal}$ is continuous with the spin in dielectric (Fig. 8) – in full compliance with the boundary conditions (11) for the electric field. However, another case of continuity is more important: the quantity

$$s_m = s^F + s_m^M = -\frac{g}{\omega}\frac{\kappa_2}{\varepsilon_2 k_s}e^{2\kappa_2 x} \tag{70}$$

is just the "naive" Minkowski spin (described by Eq. (S13) at $\tilde{\varepsilon}_2 = \varepsilon_2$). It satisfies the condition $s_m(-0) = s^F(+0)$ (see the red curves in Fig. 8), which agrees with the fact that the Minkowski spin is continuous [23] (cf. also Eq. (S13) and the note beneath (S14)). This conclusion supplies additional arguments to the correctness of Eqs. (68), and justifies the procedure chosen for the spin calculation via (56), (59) and (67) – had the spin been derived formally as (69), the Minkowski spin continuity would have been destroyed.

The "complete" spin in medium 2 (see the magenta curves in Fig. 8) is described by the equation

$$s = s^F + s_m^M + s_R^M = -\frac{g}{\omega}\frac{\kappa_2}{\varepsilon_2^2 k_s}\Big[\big(1+\eta\big)e^{2\kappa_2 x} - 2\eta e^{(\kappa_2+\gamma)x}\Big]. \tag{71}$$

Its "volume" part coincides with the phenomenological result (S13); however, the NS term in (71) provides that, in contrast to the phenomenological spin of Eq. (S13), the overall microscopic spin is continuous:

$$s(-0) = s(+0).$$

This formula replaces the phenomenological result (S14) depicted by the "step" BC: the microscopic analysis enables both the "naïve" Minkowski spin and the complete spin to be continuous at $x = 0$.

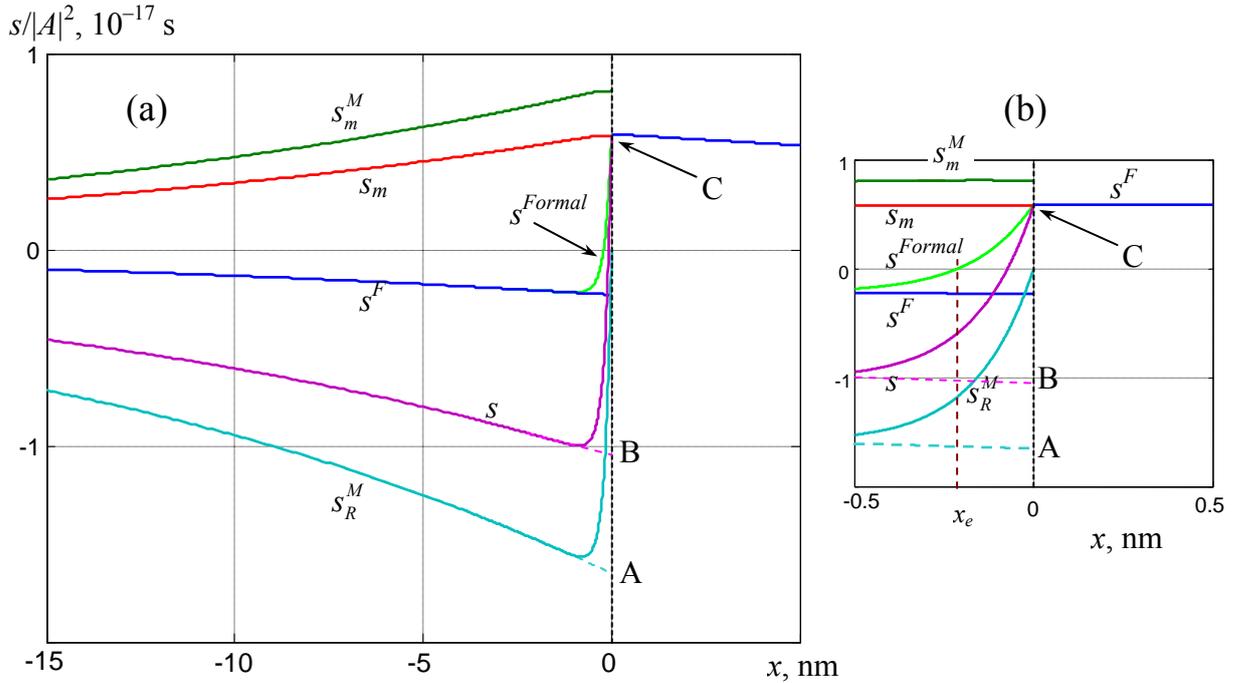

Fig. 8. SPP spin constituents calculated for the conditions (17), (18). (Blue) "field" contribution (S13) (for $x > 0$) and (68) originating from $p_S^F$ (56); (light green) "formal" "field" spin (69); (green) material contribution (68) originating from $p_{mS}^M$ (59); (red) "naïve" Minkowski spin (70); (cyan) "rotational" material spin (47), (66); (magenta) complete spin (71). Dashed lines show the phenomenologically expected behavior, the interface plane is marked by the vertical dashed line, the panel (b) shows the NS region of (a) in a magnified horizontal scale.

Another interesting observation concerns the "spin-momentum law" [36] which interprets the "extraordinary" transverse spin of the SPP through the vorticity of the longitudinal momentum density [6,36]: $\mathbf{s} \propto \nabla \times \mathbf{p}$ (see Eq. (S25)). This feature of the SPP field reveals a transparent physical analogy with the layered flow of a fluid near a plane wall, which is accompanied by the flow vorticity orthogonal to the main direction of propagation [37]. However, juxtaposition of Eqs. (71) and (50) indicates that the "spin-momentum law" (S25) is only valid for the volume contributions, while close to the interface the proportionality between the spin $\mathbf{s}$ and the curl of momentum $\mathbf{p}$ is destroyed by the NS terms.

### 4. Frequency dependence of the SPP dynamical characteristics

The results of Sections 2 and 3 provide an exhaustive description of the spatial distributions for various DCs of the SPP field. However, the illustrations were restricted to the single radiation wavelength corresponding to point A of the dispersion curve (Fig. 2). In this Section, we briefly outline the main modifications expected when the radiation frequency $\omega$ or, which is the same, the vacuum wavelength $\lambda$, changes.

We start with the remark that, according to Fig. 2, for any frequency the parameter $\gamma$ (13) remains two orders of magnitude higher than other spatial parameters $\kappa_1$ and $\kappa_2$, and all approximations associated with the condition (22) are thus valid within the wide range of possible frequencies. Consequently, all the analytical formulas of Sections 2 and 3 are appropriate, and the frequency dependence of the DCs characteristics is comprehensively determined by the frequency dependencies of the main SPP parameters $k_s$, $\kappa_1$, $\kappa_2$ presented in Fig. 2.

Fig. 9 provides a comparison between the SPP characteristics for the two situations: "high-frequency" regime specified by point B in the dispersion curve of Fig. 2 ($\omega = 5.19 \cdot 10^{15}$ s$^{-1}$, $\lambda = 363$ nm) and the "low-frequency" one (point C in Fig. 2, $\omega = 2.685 \cdot 10^{15}$ s$^{-1}$, $\lambda = 702$ nm). The first notable discrepancy between the two situations is remarkably higher absolute values of all the presented quantities in the high-frequency regime. However, this observation is not new: As it is shown in previous sections, the "volume" contributions as well as the integral values (in the sense of (21)) are correctly described by the phenomenological expressions of Section S1 (see Supplemental Document) whose frequency-dependent behavior is well known [1,2,23]. Therefore, here we focus on the NS contributions inducing some peculiar features of the DCs' spatial distributions.

For many DCs (e.g., those discussed in Figs. 5, 8), the frequency variation entails mainly quantitative transformations of the NS behavior coherent with changes in the absolute values of the "volume" terms. But more interesting NS features are associated with the transverse-electric-field sign inversion (which is rather frequency-sensitive, see the "high-frequency" $x_{eB}$ and "low-frequency" $x_{eC}$ zero-point positions in Fig. 9a, b and Fig. 10a). These are the characteristic "kinks", additional "peaks" and "troughs" in the distributions of energy and of the elements (58) − (61) of the material-momentum spin-orbital decomposition (see Figs. 4, 6, 7). The corresponding momentum constituents are presented in Fig. 9, and one can see that the shapes of these peculiar details substantially depend on the radiation frequency.

Fig. 9a shows that the "potential wells" in the NS energy distributions, distinctly seen at the "high" (and "moderate", see Fig. 4) frequencies, disappear when the frequency decreases. Interestingly, this happens because their "left walls" $\delta w_2$, $\delta w_2^F$ fall down whereas the "right walls" $\delta w_1$, $\delta w_1^F$ practically do not change. In more detail, the frequency dependence of the "left walls" is illustrated by Fig. 10a for the "field" and "complete" energy density (the corresponding data for the $w^M(x)$ can be derived from Eq. (31)).

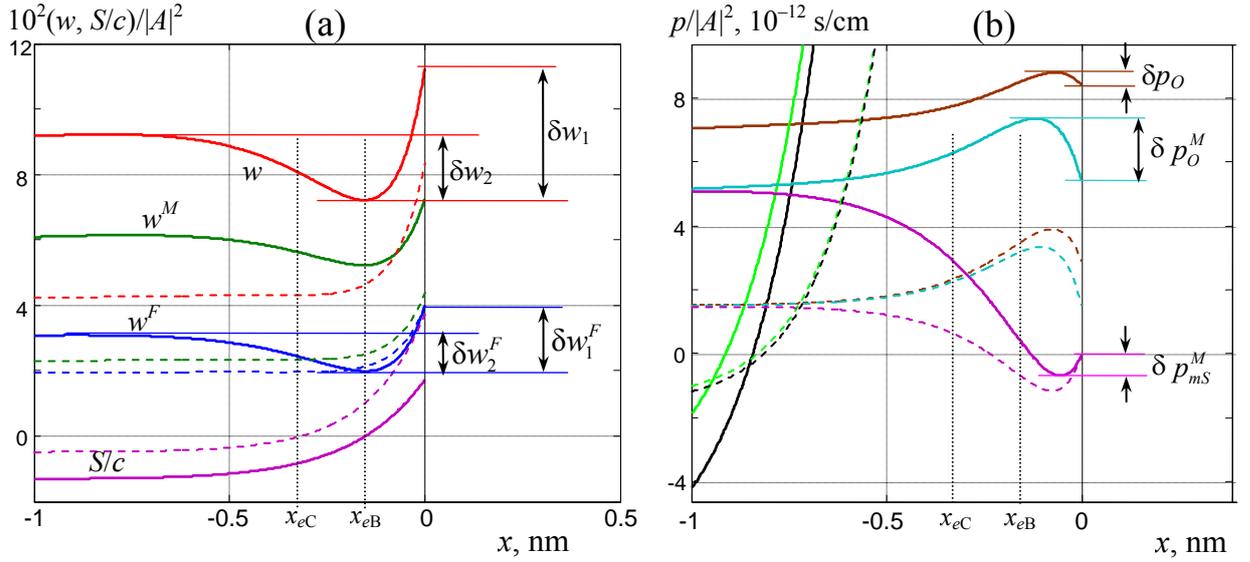

Fig. 9. Spatial distributions of the SPP DCs in medium 2 at different frequencies: (solid lines) conditions of point B in Fig. 2 ($k_s = 3.99 \cdot 10^5$ cm$^{-1}$, $k = 1.73 \cdot 10^5$ cm$^{-1}$), (dashed lines) conditions of point C in Fig. 2 ($k_s = 0.96 \cdot 10^5$ cm$^{-1}$, $k = 0.895 \cdot 10^5$ cm$^{-1}$); $x_{eB} = -0.16$ nm and $x_{eC} = -0.32$ nm are the corresponding positions where $E_x$ vanishes, see Fig. 3b. (a) Energy and energy flow density distributions (cf. Fig. 4): (blue) "field" contribution (32); (green) material contribution (33); (red) "complete" energy density (34); (magenta) energy flow density (37) and (S8). (b) SPP momentum constituents (cf. Fig. 7): (brown) complete orbital momentum (62); (black) complete spin momentum (63); (cyan) orbital material momentum (60); (light-green) complete spin material momentum (61); (magenta) spin material contribution (59). The morphology parameters of the curves $\delta w_1$, $\delta w_1^F$, $\delta w_2$, $\delta w_2^F$, $\delta p_O$, $\delta p_O^M$ and $\delta p_{mS}^M$ are characterized in more detail by Fig. 10.

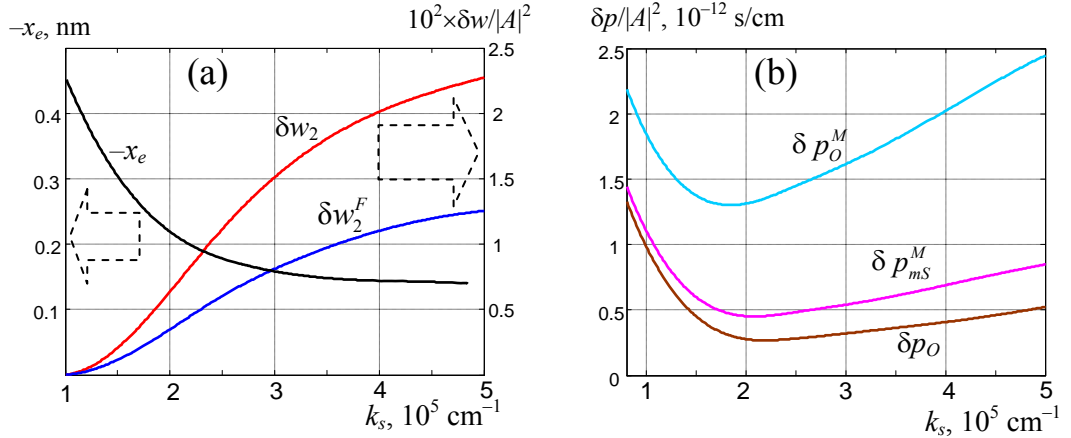

Fig. 10. Frequency dependences of the morphology parameters characterizing the NS spatial behavior of the DCs presented in Fig. 10. (a) "Depth" of the energy "wells" of Fig. 10a (red and blue, right vertical scale) and positions $x_e$ at which $E_x = 0$ (black, left vertical scale); (b) "peak heights" and "trough depth" of the momentum constituents' distributions of Fig. 9b; the curve colors are the same as colors of the corresponding curves in Fig 9b.

Similarly, Fig. 9b juxtaposes the "low-frequency" and "high-frequency" "peaks" and "troughs" of $p_O$, $p_O^M$ and $p_{mS}^M$ distributions (cf. Fig. 7). In contrast to Fig. 9a, where the "well" centers (energy minima) move "inwards" the medium 2 together with $x_e$ (cf. the black curve in Fig. 10a), the extremum positions of the "kinks" are less frequency-sensitive and always lie closer to the interface. However, the "kinks" morphology characterized by the parameters $\delta p_O$, $\delta p_O^M$ and $\delta p_{mS}^M$ explained in Fig. 9b shows more interesting frequency-dependent evolutions presented in Fig. 10b.

The data of Figs. 9 and 10 supply a general characterization of the frequency-induced changes in the NS spatial distributions of the selected DCs. In particular, they show possible ways for the deliberate control of these spatial distributions, especially, for purposeful formation of the "kinks", "peaks" and "troughs" of the corresponding curves inside the nanometer-scale vicinity of the interface.

## 5. Concluding remarks

The main outcome of this paper is the detailed analysis of the near-interface spatial behavior of the DCs in a lossless SPP wave. Due to the use of the hydrodynamic model, extremely subwavelength structures are revealed in the distributions of the field energy and momentum. Importantly, the paper presents the first to our knowledge collection of analytical expressions describing the separate constituents of the electromagnetic momentum in plasmonic fields, with the explicit identification of the "field" and material contributions of different origins. For a consistent comparison with the known phenomenological results, which are expressed in terms of the spin-orbital (canonical) momentum decomposition, the procedure of such decomposition in presence of free charges and currents is developed. The procedure validity is substantiated by using the SPP fields as a model example.

Within the limits of common applicability, the new details of the DC spatial distributions comply with the known SPP properties described by the phenomenological approach and/or by the simplified Drude model of the metallic medium. However, the new results demonstrate the existence of controllable nano-scale optical structures in the NS layer of the medium 2 that can be deliberately created and whose parameters can be efficiently regulated, e.g., by choosing the SPP frequency.

Although the present consideration deals with the homogeneous metallic medium 2, it can be generalized to more complex plasmonic structures containing the metamaterials, composite media, fluids, nanoclusters, etc. [39–42]. One can expect that specific local interactions between the field and material nano-objects (e.g., impurity atoms, molecules, lattice defects and nanoclusters introduced into the metal matrix [43]) can be inspired and mediated by the nanoscale field inhomogeneities described in the paper, and the degree of spatial commensurability between the field "kinks" and the nano-objects supplies additional channel for controlling this interaction. The subwavelength field structures can be useful in various problems of nano-optics: for example, the "peaks" and "troughs" in the NS energy distribution (Figs. 4, 9a) can serve for precise localization of the objects, their spatially-selective excitation or transformation, etc. Usually, such nano-objects are characterized by a noticeable mobility in the crystal structure, or the medium 2 can be liquid [44]; then, the controllable "kinks" in the momentum distributions (Figs. 6, 7, 9b) may be employed for their guiding and transportation due to the field's mechanical action [38].

Especially interesting manifestations of the specific DC-distribution details predicted in this paper are expected in combination with the external or SPP-induced static fields, for example, the SPP-induced magnetization [20–23], which is proportional to the material spin constituent $s_R^M$ (66). In particular, the nanoscale variations of the energy and momentum distributions may affect the electron spin transport processes [45–47], induce additional influences on the photo-sensitive centers in presence of high-gradient local optical fields [48], etc. Despite the presence of a non-zero spin density, which is a hallmark of SPP fields [7,8,11,12], the usual TM SPPs produce no specific chiral action because their transverse spin is helicity-independent. However, the use of specific chiral metasurfaces [49,50] enables generation of chiral SPPs with controllable spin orientation, which can selectively excite and control chiral nano-objects, e.g. the diamond spin-qubits [51].

In conclusion, let us outline the main restrictions of the presented approach and possible ways of its further development. The first limitation originates from the lossless approximation, which is widely used in the SPP theory [2–8,12,13,20,23] and enables the physically transparent analytical description. In principle, the effects of damping (energy dissipation) can be taken into account by an additional summand $-mn_0\Gamma\mathbf{v}$ in the right-hand side of Eq. (10) where $\Gamma$ is the collision frequency [30,32]. After this modification, the formal solution of Eqs. (1), (9), (10) and (11) can be derived directly from Eqs. (12) − (16) via replacements

$$m \to m\left(1 + i\frac{\Gamma}{\omega}\right), \quad \beta^2 \to \beta^2 \frac{1}{1 + i\Gamma/\omega},$$

which, in turn, lead to the modified expressions (7) and (13),

$$\eta \to \frac{\omega_p^2}{\omega^2}\frac{1}{1 + i\Gamma/\omega}, \quad \gamma^2 = k_s^2 - \frac{\omega^2\varepsilon_2}{\beta^2}\left(1 + i\frac{\Gamma}{\omega}\right). \tag{72}$$

As a result, the expression for $\varepsilon_2$ (7) is transformed to the usual complex permittivity expression for the Drude model with losses [2], and the constant $\gamma$ becomes complex. Unfortunately, such direct modifications cannot be applied to the DC expressions because these are quadratic in field and electron velocity components and include the real and imaginary parts separately; however, some general conclusions can be inferred from the simple reasoning. In order the propagating SPP to exist, the damping correction must be small; for example, in the considered case of silver, $\Gamma = 5.9\cdot10^{13}$ s$^{-1}$ [52] so for the conditions of points B, A and C of Fig. 2, $\Gamma/\omega = 0.049$, 0.014 and 0.022, respectively. Such small corrections mean that in the most interesting cases, the loss-induced modifications of the results are mainly quantitative. Nevertheless, some corrections are principal. For example, due to complex $\eta$ and $\gamma$ (72), the electric field component $E_x$ (12) differs from zero everywhere in the medium 2: while Re($E_x$) vanishes at $x = x_e$, the small "imaginary" field $E_x$, which is nearly in-phase with the longitudinal component $E_z$, still exists. Accordingly, the singularity in $\alpha_{xx}$ (27) is removed. Practically this means that the "kinks" in the energy and momentum distributions become smoothened, as well as the "singular" NS terms in Eqs. (16), (48) − (50). Anyway, the detailed analysis of the DC behavior in presence of energy losses deserves a special investigation in a separate work.

The second important simplification is the assumption of perfectly flat interface, which is frequently broken in practice [53–55]. Usually the influences of roughness are evaluated by statistical methods but, for the effects considered here, this way of reasoning cannot be applied because even very small boundary imperfections of the order of $|x_e|$ (see Figs. 3 − 8) are crucial. Most probably, the irregularities of such size scale will induce local SPP-field confinement and enhancements [29] but their correct analysis can hardly be performed analytically and requires exquisite numerical approaches (e.g. the finite-element method with non-uniform meshing [29]). One may expect that the out-of-plane boundary deviations of the size ~$|x_e|$ will not destroy the predicted pictures as far as their size in the ($y$–$z$)-plane (Fig. 1) exceeds $|x_e|$ by the order of magnitude. Alternatively, the interface irregularities can be used for purposeful formation of "hot spots" in the SPP fields [53].

The concept of the "interface imperfections" makes a natural "bridge" to the complex SPP-supporting structures where the media 1 and 2 are separated by the specially created metasurfaces, contributing, for example, to the formation of chiral SPPs [49,50]. Usually, such meta-boundaries form the physically selected nanometer-wide intermediate layer between the metal and dielectric media, which in calculations is replaced by the conventional plane interface with specific electromagnetic properties [50]. The applicability of the existing hydrodynamic model in the vicinity of this conventional interface is questionable but its further generalization to such situations looks promising and relevant.

And the last approximation, which deserves a few remarks, is the linearity of the basic equations (1) and (9) − (11). In contrast to the usual SPP theory where the linear character is accepted as needless to say, in this paper the situations of extraordinary high local values of some parameters occurs systematically, e.g., in connection with the "singular" NS terms in Eqs. (16), (48) − (50). This requires a special attention: for example, if in Eq. (20) the non-equilibrium concentration of electrons $n(0)$ exceeds or even becomes comparable with the "background" concentration $n_0$, this is physically inappropriate. Estimations made in Sec. 2.2 show that the requirements to the admissible field intensities are rather loose. If we consider an SPP wave with the total power $Q$ propagating within the width $\Delta y$, the electric field amplitude can be determined via Eq. (S10) as

$$|A|^2 = \frac{16\pi k_s \kappa_2 \varepsilon_2 \varepsilon_1^2}{\omega\left(\varepsilon_2^2 - \varepsilon_1^2\right)\Delta y} Q \,,$$

which yields that even for $Q = 1$ W and $\Delta y = 100$ μm (other parameters can be taken from Eqs. (17) − (19)), $|A| \approx 500$ statV/cm, which, in view of the note below Eq. (20), means $n(0)/n_0 \approx 3.7 \cdot 10^{-4}$. This value, probably, warrants the model applicability in many practical situations; however, in higher fields which can easily occur due to microscopic dimensions of the system [29], the improvements accounting for the quantum character of the electrons' behavior, including the quantum tunneling effects [29] as well as the non-linear response of the media [4,56], would be necessary.

**Funding.** Ministry of Education and Science of Ukraine (582/18), Research Institute of Zhejiang University – Taizhou, Center for Modern Optical Technology, China.

**Acknowledgments.** The authors are grateful to Konstantin Bliokh (RIKEN, Japan) for the stimulating discussions.

**Disclosures.** The authors declare no conflicts of interest.

**Data availability.** No data were generated or analyzed in the presented research.

**Supplemental Document.** See Supplement 1 for supporting content.

# Microscopic analysis of the energy, momentum and spin distributions in a surface plasmon-polariton wave

**Supplemental Document**

### S1. Phenomenological characterization of the SPP field

S1.1. Energy distribution.

Taking into account dispersion of the medium, the energy density is determined by the Brillouin's formula [24,25]

$$w = \frac{1}{16\pi}\left(\tilde{\varepsilon}|\mathbf{E}|^2 + \tilde{\mu}|\mathbf{H}|^2\right) = \frac{1}{16\pi}\left[\tilde{\varepsilon}\left(|E_x|^2 + |E_z|^2\right) + |H_y|^2\right] \tag{S1}$$

where $\mu = \tilde{\mu} = 1$ (the medium is non-magnetic) and

$$\tilde{\varepsilon} = \frac{d}{d\omega}\left[\omega\varepsilon(\omega)\right], \tag{S2}$$

giving (see Eq. (7))

$$\tilde{\varepsilon}_1 = \varepsilon_1, \quad \tilde{\varepsilon}_2 = 1 + \eta = 2 - \varepsilon_2. \tag{S3}$$

Accordingly,

$$w = g\begin{cases} \dfrac{1}{\varepsilon_1}e^{-2\kappa_1 x}, & x > 0; \\[2mm] \dfrac{1}{\varepsilon_2^2}\left(1 + \dfrac{\varepsilon_2^2 - \varepsilon_2}{\varepsilon_1}\right)e^{2\kappa_2 x}, & x < 0. \end{cases} \tag{S4}$$

where $g = |A|^2/(8\pi)$. This distribution shows a discontinuity at $x = 0$

$$\Delta w = w(+0) - w(-0) = g\frac{1}{\varepsilon_2}\left(\frac{1}{\varepsilon_1} - \frac{1}{\varepsilon_2}\right). \tag{S5}$$

The total energy of the SPP (in the sense of Eq. (21)) equals to

$$\langle w\rangle = \int_{-\infty}^{\infty} w\,dx = \frac{g}{2}\left(\frac{1}{\kappa_1} + \frac{1}{\kappa_2}\right)\left(\frac{1}{\varepsilon_1} + \frac{1}{\varepsilon_2^2}\right) = \frac{g}{2\kappa_2}\left(1 - \frac{\varepsilon_2}{\varepsilon_1}\right)\left(\frac{1}{\varepsilon_1} + \frac{1}{\varepsilon_2^2}\right). \tag{S6}$$

S1.2. Electromagnetic energy flow density is meaningfully described by the kinetic Abraham momentum [3,7,20,21]:

$$\mathbf{S} = \frac{c}{8\pi}\text{Re}\left[\mathbf{E}^* \times \mathbf{H}\right] = \frac{c}{8\pi}\mathbf{z}E_x^* H_y. \tag{S7}$$

With employment of Eqs. (3) – (5), Eq. (S7) means

$$\mathbf{S} = \mathbf{z}S, \quad S = g\frac{\omega}{k_s}\begin{cases} \dfrac{1}{\varepsilon_1}e^{-2\kappa_1 x}, & x > 0; \\[2mm] \dfrac{1}{\varepsilon_2}e^{2\kappa_2 x}, & x < 0. \end{cases} \tag{S8}$$

The energy flow is positive in medium 1 and negative in medium 2; the transition occurs discontinuously at $x = 0$ with the step

$$\Delta S = S(+0) - S(-0) = g\frac{\omega}{k_s}\left(\frac{1}{\varepsilon_1} - \frac{1}{\varepsilon_2}\right). \tag{S9}$$

The total energy flow in the $z$-direction per unit $y$-width is easily obtained from Eqs. (21), (S8) and (5),

$$\langle S \rangle = g \, \frac{\omega}{2k_s} \frac{1}{\varepsilon_1 \kappa_1} \left( 1 - \frac{\varepsilon_1^2}{\varepsilon_2^2} \right) = -g \, \frac{\omega}{2k_s} \frac{1}{\varepsilon_2 \kappa_2} \left( \frac{\varepsilon_2^2}{\varepsilon_1^2} - 1 \right), \tag{S10}$$

which enables us to calculate the group velocity of the whole SPP wave packet $\mathbf{v}_g = \mathbf{z} v_g$,

$$v_g = \frac{\langle S \rangle}{\langle w \rangle} = \frac{\omega}{k_s} \frac{\varepsilon_2 \left( \varepsilon_1 + \varepsilon_2 \right)}{\varepsilon_1 + \varepsilon_2^2}. \tag{S11}$$

Note that due to condition (6) the total energy flow $\langle S \rangle$ and the group velocity $v_g$ are positive.

## S1.3. Spin density

The spin density in a dispersive medium is determined by the formula that follows from the Minkowski spin via the same substitution (S2), (S3) as is made in the Brillouin formula (S1) for the energy [18–22]:

$$\mathbf{s} = \frac{1}{16\pi\omega} \mathrm{Im} \left( \tilde{\varepsilon} \mathbf{E}^* \times \mathbf{E} + \tilde{\mu} \mathbf{H}^* \times \mathbf{H} \right) = \frac{\tilde{\varepsilon}}{16\pi\omega} \mathbf{y} \, \mathrm{Im} \left( E_z^* E_x - E_x^* E_z \right). \tag{S12}$$

Being applied to Eqs. (3) and (4), Eq. (S12) gives $\mathbf{s} = \mathbf{y} s$,

$$s = \frac{g}{\omega k_s} \begin{cases} \dfrac{\kappa_1}{\varepsilon_1} e^{-2\kappa_1 x}, & x > 0; \\ -\dfrac{\tilde{\varepsilon}_2}{\varepsilon_2} \dfrac{\kappa_2}{\varepsilon_2} e^{2\kappa_2 x} = \left( 1 - \dfrac{2}{\varepsilon_2} \right) \dfrac{\kappa_2}{\varepsilon_2} e^{2\kappa_2 x}, & x < 0. \end{cases} \tag{S13}$$

Typically for the evanescent waves [7–12], the spin is directed orthogonally to the SPP propagation (red arrow in Fig. 1), and the spin directions in both media are opposite in agreement with the opposite directions of the $\mathbf{E}$ vector rotation in media 1 and 2. The SPP spin (S13) is, generally, discontinuous at the interface $x = 0$:

$$\Delta s = s \left( +0 \right) - s \left( -0 \right) = \frac{g}{\omega k_s} \frac{\kappa_1}{\varepsilon_1} \left( 1 - \frac{\tilde{\varepsilon}_2}{\varepsilon_2} \right) = \frac{2g}{\omega k_s} \frac{\kappa_1}{\varepsilon_1} \left( 1 - \frac{1}{\varepsilon_2} \right). \tag{S14}$$

This discontinuity owes completely to the dispersion [23]: when $\tilde{\varepsilon}_2 = \varepsilon_2$, Eqs. (S13), (S14) describe the "naïve" (dispersion-free) Minkowski spin, which is continuous due to the definition (S12) and the boundary conditions (2).

The total spin of the SPP (in the sense of Eq. (21)) is determined via the integration of the spin density (S13) over the whole range of $x$, which, with account for Eqs. (S10) and (S6), gives

$$\langle s \rangle = \frac{g}{2\omega k_s \varepsilon_1} \left( 1 - \frac{\tilde{\varepsilon}_2}{\varepsilon_2} \frac{\varepsilon_1}{\varepsilon_2} \right) = \frac{g}{2\omega k_s \varepsilon_1} \left( 1 + \frac{\varepsilon_1}{\varepsilon_2} - \frac{2\varepsilon_1}{\varepsilon_2 \varepsilon_2} \right). \tag{S15}$$

## S1.4. Orbital momentum density

Another DC is the orbital (canonical) momentum [20,21],

$$\mathbf{p}_O = \frac{1}{16\pi\omega} \mathrm{Im} \left[ \tilde{\varepsilon} \mathbf{E}^* \cdot (\nabla) \mathbf{E} + \tilde{\mu} \mathbf{H}^* \cdot (\nabla) \mathbf{H} \right] = \mathbf{z} \frac{k_s}{16\pi\omega} \left[ \tilde{\varepsilon} \left( |E_x|^2 + |E_z|^2 \right) + |H_y|^2 \right]. \tag{S16}$$

The second equality (S16) results from application of the first one to the SPP field (3), (4). Comparing Eq. (S16) with Eq. (S1) immediately reveals that the spatial distribution of the orbital momentum is proportional to the energy distribution (S1) – (S6):

$$\mathbf{p}_O = \mathbf{z} p_O, \quad p_O = \frac{k_s}{\omega} w, \tag{S17}$$

or

$$p_O = \frac{g}{\omega} k_s \begin{cases} \dfrac{1}{\varepsilon_1} e^{-2\kappa_1 x}, & x > 0; \\[2mm] \dfrac{1}{\varepsilon_2^2}\left(1 + \dfrac{\varepsilon_2^2 - \varepsilon_2}{\varepsilon_1}\right) e^{2\kappa_2 x}, & x < 0. \end{cases} \tag{S18}$$

The integral orbital momentum follows immediately from (S17) and (S6),

$$\langle p_O \rangle = \frac{g k_s}{2\omega \kappa_2}\left(1 - \frac{\varepsilon_2}{\varepsilon_1}\right)\left(\frac{1}{\varepsilon_1} + \frac{1}{\varepsilon_2^2}\right). \tag{S19}$$

## S1.5. Spin momentum density

The spin momentum (Belinfante momentum) [7,8,27,28] of the electromagnetic field is directly related with the spin (S12) [22]:

$$\mathbf{p}_S = \frac{1}{2}\nabla \times \mathbf{s} = \frac{1}{2}\left(-\mathbf{x}\frac{\partial s_y}{\partial z} + \mathbf{z}\frac{\partial s_y}{\partial x}\right), \quad \mathbf{p}_S = \mathbf{z}p_S, \tag{S20}$$

(the second equality (S20) follows from the first one applied to the SPP field). Taking Eq. (S13) into account we obtain

$$p_S = \frac{g}{\omega}k_s \begin{cases} \dfrac{1}{\varepsilon_2}e^{-2\kappa_1 x}, & x > 0 \\[2mm] \dfrac{2-\varepsilon_2}{\varepsilon_1\varepsilon_2}e^{2\kappa_2 x}, & x < 0 \end{cases} + p^{\mathrm{surf}}, \quad p^{\mathrm{surf}} = \frac{g}{2\omega k_s}\frac{\kappa_2}{\varepsilon_2}\left(\frac{\tilde{\varepsilon}_2}{\varepsilon_2}-1\right)\delta(x) = \frac{g}{\omega}\frac{k_s}{\varepsilon_2^2\kappa_1}\eta\delta(x). \tag{S21}$$

The spin momentum consists of the volume part (in curly brackets) and the surface contribution described by the delta-function, which formally follows from the spin discontinuity (S14). This singular term is due to the dispersion and vanishes if $\tilde{\varepsilon}_2$ is replaced by $\varepsilon_2$. Importantly, the volume and surface contributions of the spin momentum exactly compensate each other so that the total spin momentum integrated over the whole space vanishes, as is required by the general theory [27,28]:

$$\langle p_S \rangle = \int_{-\infty}^{\infty} p_S \, dx = 0. \tag{S22}$$

Equations (S18) and (S21) enable us to find the "overall" momentum density, which is represented by the sum of the spin and orbital contributions,

$$p = p_S + p_O = \frac{g}{c}\frac{k}{k_s}\left[\begin{cases} e^{-2\kappa_1 x}, & x > 0 \\[2mm] \dfrac{1}{\varepsilon_2}e^{2\kappa_2 x}, & x < 0 \end{cases} + \frac{\kappa_2}{\varepsilon_2 k^2}\left(\frac{1}{\varepsilon_2}-1\right)\delta(x)\right]. \tag{S23}$$

In view of (S17) and (S22), the total momentum of SPP acquires the form

$$\langle p \rangle = \langle p_O \rangle = \frac{k_s}{\omega}\langle w \rangle = \sqrt{\frac{\varepsilon_1\varepsilon_2}{\varepsilon_1 + \varepsilon_2}}\frac{k}{\omega}\langle w \rangle. \tag{S24}$$

Generally (while $\varepsilon_1^2/(1-\varepsilon_1) < \varepsilon_2 < -\varepsilon_1$), the local SPP momentum (S24) exceeds the momentum of a plane wave with the same energy $\varepsilon_1 k/\omega\langle w\rangle$ [7,8].

Confrontation of Eqs. (S23) and (S13) yields the "spin-momentum law" for the SPP wave [36] that can be presented in the form

$$\mathbf{s} = \mathbf{y}s = \begin{cases} \dfrac{1}{2k_1^2} \nabla \times \mathbf{p}, & x > 0; \\ \dfrac{1}{2k_2^2} \dfrac{2\tilde{\varepsilon}_2}{\tilde{\varepsilon}_2 + \varepsilon_2} \nabla \times \mathbf{p} = \dfrac{2-\varepsilon_2}{2k_2^2} \nabla \times \mathbf{p}, & x < 0. \end{cases} \tag{S25}$$

where $\nabla \times \mathbf{p} = \nabla \times (\mathbf{z}p) = -\mathbf{y}\dfrac{dp}{dx}$, $k_i^2 = k^2 \varepsilon_i$ $(i = 1, 2)$, and only the volume part of the momentum is considered (the $\delta$-term of (S23) is excluded).

### S2. Derivation of the material momentum contributions

In an ideal monochromatic field all average processes should be stationary, and the controversy appears: either the constant force produces the variable momentum or the stationary momentum implies the zero force. To avoid this controversy and following [16], we temporarily consider a quasi-monochromatic field where the complex electric-field phasor, introduced in the 1st paragraph of Section 2.1, may slowly change with time, $\mathbf{E} \to \mathbf{E}(t)$ (wave packet). This means that

$$\mathbf{E}(t) = \int_{-\infty}^{\infty} \tilde{\mathbf{E}}(\theta) e^{-i\theta t} \frac{d\theta}{2\pi} \tag{S26}$$

where the Fourier amplitude $\tilde{\mathbf{E}}(\theta)$ differs noticeably from zero only in the narrow interval $|\theta| << \omega$. Further, for the monochromatic case, the linear dependence (26) takes place, which for the slowly varying field (S26) leads to

$$\mathbf{d}(t) = \int_{-\infty}^{\infty} \hat{\alpha}(\omega + \theta) \tilde{\mathbf{E}}(\theta) e^{-i\theta t} \frac{d\theta}{2\pi}.$$

Since the spectral width of the field (S26) can be arbitrary small, the approximation

$$\hat{\alpha}(\omega + \theta) = \hat{\alpha}(\omega) + \theta\hat{\alpha}'(\omega), \quad \hat{\alpha}'(\omega) = \frac{d\hat{\alpha}(\omega)}{d\omega}$$

is valid, and

$$\mathbf{d}(t) = \hat{\alpha}(\omega)\mathbf{E}(t) + i\hat{\alpha}'(\omega)\frac{\partial}{\partial t}\mathbf{E}(t). \tag{S27}$$

To find the force, we start with the standard expression for the dipole force $\mathbf{F}(t)$ involving the real time-dependent field and dipole characteristics $\mathbf{E}(t)$, $\mathbf{H}(t)$ and $\mathbf{d}(t)$ but slightly modify its usual form using the Maxwell equations in vacuum [24,25]:

$$\mathbf{F} = (\mathbf{d} \cdot \nabla)\mathbf{E} + \frac{1}{c}\frac{\partial \mathbf{d}}{\partial t} \times \mathbf{H} = (\mathbf{d} \cdot \nabla)\mathbf{E} + \mathbf{d} \times (\nabla \times \mathbf{E}) + \frac{1}{c}\frac{\partial}{\partial t}(\mathbf{d} \times \mathbf{H}).$$

Now this expression can be averaged over the oscillation period introducing slowly varying phasors $\mathbf{F}(t)$, $\mathbf{d}(t)$, etc., via the typical substitutions

$$\mathbf{F}(t) \to \frac{1}{2}\left[\mathbf{F}(t)e^{-i\omega t} + \mathbf{F}^*(t)e^{i\omega t}\right], \quad \mathbf{d}(t) \to \frac{1}{2}\left[\mathbf{d}(t)e^{-i\omega t} + \mathbf{d}^*(t)e^{i\omega t}\right], \dots$$

which results in

$$\mathbf{F} = \frac{1}{2}\text{Re}\left[(\mathbf{d}^* \cdot \nabla)\mathbf{E} + \mathbf{d}^* \times (\nabla \times \mathbf{E}) + \frac{1}{c}\frac{\partial}{\partial t}(\mathbf{d}^* \times \mathbf{H})\right]. \tag{S28}$$

This form is more suitable for the momentum calculation. First, the last summand explicitly exposes one of the momentum contributions,

$$\mathbf{p}_m^M = \frac{1}{2c}\text{Re}(\mathbf{d}^* \times \mathbf{H}) = \frac{1}{2c}\text{Re}(\hat{\alpha}\mathbf{E}^* \times \mathbf{H}), \tag{S29}$$

which, with the help of Eqs. (10) and (25), can be presented as Eq. (41).

The other terms of (S28) can be transformed in the similar way, with allowance for (S27):

$$\mathbf{F} = \frac{1}{2}\mathrm{Re}\left[\left(\mathbf{d}^* \cdot \nabla\right)\mathbf{E} + \mathbf{d}^* \times \left(\nabla \times \mathbf{E}\right)\right] = \frac{1}{4}\left(\alpha_{xx}\nabla\left|E_x\right|^2 + \alpha_{zz}\nabla\left|E_z\right|^2\right)$$

$$+ \frac{1}{2}\mathrm{Re}\left[\mathbf{x}\left(-i\alpha'_{xx}\frac{\partial E_x^*}{\partial t}\frac{\partial E_x}{\partial x} - i\alpha'_{zz}\frac{\partial E_z^*}{\partial t}\frac{\partial E_z}{\partial x}\right) + \mathbf{z}\left(-i\alpha'_{xx}\frac{\partial E_x^*}{\partial t}\frac{\partial E_x}{\partial z} - i\alpha'_{zz}\frac{\partial E_z^*}{\partial t}\frac{\partial E_z}{\partial z}\right)\right]. \qquad (S30)$$

In the monochromatic limit (the wave packet length tends to infinity), $z$-derivatives in the first line vanish and the corresponding contribution reduces to the gradient force [16]

$$\mathbf{F}^G = \frac{1}{4}\mathbf{x}\left(\alpha_{xx}\frac{\partial}{\partial x}\left|E_x\right|^2 + \alpha_{zz}\frac{\partial}{\partial x}\left|E_z\right|^2\right).$$

In the case of mechanical equilibrium, this force is cancelled by a pressure gradient and can be ignored in our present context [16]. At the same time, since, due to Eqs. (12),

$$\frac{\partial E_x^*}{\partial t}\frac{\partial E_x}{\partial x} \quad \text{and} \quad \frac{\partial E_z^*}{\partial t}\frac{\partial E_z}{\partial x}$$

are real quantities, the first summand of the second line in (S30) gives zero contribution while the second summand, in view of

$$\frac{\partial E_x}{\partial z} = ik_s E_{2x}, \quad \frac{\partial E_z}{\partial z} = ik_s E_{2z},$$

can be written as $\partial \mathbf{p}_e^M / \partial t$ where $\mathbf{p}_e^M$ is determined by Eq. (44).